\documentclass[%
reprint,
aps, prb
]{revtex4-1}
\usepackage{amsmath,amssymb}
\usepackage{graphicx}
\usepackage{dcolumn}
\usepackage{bm}
\usepackage{textcomp}
\usepackage{lipsum}
\usepackage{physics}
\usepackage{hyperref}
\usepackage{xcolor}
\hypersetup{colorlinks=true, allcolors=blue}
\usepackage{enumitem}
\usepackage{multirow}
\newlist{symb}{itemize}{1}
\setlist[symb]{label=,labelwidth=0.5in,align=parleft,,leftmargin=!}

\begin{document}

\title{Dynamics of Magnetoelectric Reversal of Antiferromagnetic Domain}


\author{Arun Parthasarathy}
\email[Email: ]{arun.parth@nyu.edu}
\affiliation{\text{Electrical and Computer Engineering, New York University, Brooklyn, New York 11201, USA}}

\author{Shaloo Rakheja}
\affiliation{\text{Electrical and Computer Engineering, New York University, Brooklyn, New York 11201, USA}}


\date{\today}

\begin{abstract}
When electric and magnetic fields are applied together on a magnetoelectric antiferromagnet, the domain state is subject to reversal. Although the initial and final conditions are saturated single-domain states, the process of reversal may decompose into local multi-domain switching events. In thin films of Cr\textsubscript{2}O\textsubscript{3}, the magnetoelectric coercivity and the switching speed found from experiments are considerably lower than expected from magnetic anisotropy, similar to Brown's paradox in ferromagnetic materials. Multi-domain effects originate because antiferromagnetic domain walls are metastably pinned by lattice defects, not due to reduction of magnetostatic energy, which is negligible. This paper theoretically analyzes domain reversal in thin-film magnetoelectric antiferromagnets in the form of nucleation, domain wall propagation, and coherent rotation. The timescales of reversal mechanisms are modeled as a function of applied magnetoelectric pressure. The theory is assessed with reference to latest experimental works on magnetoelectric switching of thin-film Cr\textsubscript{2}O\textsubscript{3}: domain wall propagation is found to be dominant and responsible for switching in the experiments. The results bear implications in the energy-delay performance of ME memory devices utilizing antiferromagnetic insulators, which are prospective for  nonvolatile technology.
\end{abstract}

\pacs{}

\maketitle

\section{Introduction}

For storing digital information on a magnetic medium, the single-domain states are of interest because they can be switched reversibly and read easily. Magnetoelectric (ME) coupling establishes a direct way to realize voltage control of the domain state in antiferromagnetic insulators which have combined space and time inversion symmetry.~\cite{fiebig2005revival} This is promising for implementing compact, nonvolatile memory and logic technology with ultralow power consumption.~\cite{manipatruni2018scalable, dowben2018towards} In a recent work of interest,~\cite{kosub2017purely} a purely ME memory cell has been experimentally demonstrated using the ME antiferromagnet Cr\textsubscript{2}O\textsubscript{3}, which can be isothermally switched and read all-electrically at room temperature. See Appendix~\ref{Sec:timeline} for a brief timeline of the research on ME effect in Cr\textsubscript{2}O\textsubscript{3}. There is plenty of room for studies involving ME switching dynamics and its understanding particularly in thin-film systems, which are fundamental for spintronic applications.

When electric and magnetic fields are simultaneously applied on a ME antiferromagnet, the domain state is subject to reversal.~\cite{martin1966antiferromagnetic} Although the initial and final conditions are saturated single-domain states, the switching transient can exhibit multi-domain characteristics. Unlike in ferromagnets, the stability of multi-domain states in antiferromagnets is not apparent because the magnetostatic energy is insignificant compared to the energy of domain wall (see Appendix~\ref{Sec:issues}). More than 60 years ago, Li~\cite{li1956domain} theorized that at the N\'eel temperature $T_\text{N}$, local nucleations of the antiferromagnetic order and their subsequent growth would result in the formation of domain walls, if the wall energy is offset by the gain in entropy. Below $T_\text{N}$, the domain wall would owe its stability to the presence of lattice defects. 

Indeed, experiments have revealed multidomain structures in Cr\textsubscript{2}O\textsubscript{3} after zero-field cooling through $T_\text{N}$, both in the bulk~\cite{fiebig1995domain} and on the surface of thin film.~\cite{wu2011imaging} The bulk sample shows ``temperature memory effect"~\cite{brown1969magneto}: the domain structure does not change when the sample is heated above and cooled below $T_\text{N}$. For the thin film sample, the surface exhibits height variations of a few nanometers over a micrometer lateral scale; the typical domain size also happens to be about a micrometer. These observations reinforce the notion that defects can metastably pin domain walls and cause multi-domain switching in antiferromagnets.

In a recent study of ME switching of exchange bias in thin-film Cr\textsubscript{2}O\textsubscript{3}/Co/Pt,~\cite{toyoki2015magnetoelectric} the threshold ME pressure for reversal and the switching speed are found to be considerably lower than expected from magnetic anisotropy, similar to Brown's paradox in ferromagnetic materials.~\cite{shtrikman1960resolution} This indicates that domain reversal is decomposing into local multi-domain switching processes, instead of undergoing uniform rotation. 

In this paper, we analyze the mechanisms that govern domain reversal in thin-film ME antiferromagnets and model their timescales as a function of applied ME pressure (Sec.~\ref{Sec:theory}). The theory primarily requires the knowledge of molecular fields of the material at a given temperature. We assess the theory with respect to recent experimental works~\cite{toyoki2015magnetoelectric, kosub2017purely} on ME switching of thin-film Cr\textsubscript{2}O\textsubscript{3} (thickness $\leq$ 500 nm~\cite{ashida2014observation}) to characterize the switching dynamics (Sec.~\ref{Sec:appl}). The dominant reversal mechanism in the experiments is identified and the switching speed has been estimated accordingly. The results provide insights into the energy-delay prospects of ME memory devices.

\section{Background}
\subsection{Linear ME effect in \texorpdfstring{C\MakeLowercase{r}\textsubscript{2}O\textsubscript{3}}{Cr2O3}}

The linear ME effect in the presence of electric field $\vb{E}$ and magnetic field $\vb{H}$ is given by the free energy density $\mathcal{F}_\mathrm{ME} = -\alpha_{ij} {E}_i{H}_j$, where $\alpha_{ij}$ is the ME tensor.~\cite{fiebig2005revival} The symmetry constraints in Cr\textsubscript{2}O\textsubscript{3} allow non-zero ME effect only when the electric and magnetic fields are parallel.~\cite{dzyaloshinskii1960magneto} This reduces the independent components of the ME tensor to $\alpha_\parallel$ and $\alpha_\perp$, for the directions parallel and perpendicular to the trigonal $z$-axis,~\cite{mostovoy2010temperature} respectively. 

In practice, the fields are applied exclusively along the $c$-axis because $\alpha_\parallel \sim 1$ ps/m~\cite{borisov2007superconducting} is an order of magnitude larger than $\alpha_\perp$ at room temperature,~\cite{folen1961anisotropy} and the uncompensated magnetization of the (0001) surface allows for easier detection.~\cite{wu2011imaging} Hence, the ME energy density can be simplified to 
\begin{equation}
\mathcal{F}_\mathrm{ME} =  -\alpha_{\parallel}{EH} \cos\theta,
\label{eq:MEenergy} 
\end{equation}
where $\theta$ denotes the direction of the antiferromagnetic order parameter relative to the $z$-axis. The signs of $E$ and $ H$ are determined according to orientation of the $z$-axis. For a 180$^{\circ}$ domain wall, the ME effect creates a pressure difference $\mathcal{F} = \abs{2\alpha_{\parallel}{EH}}$ across the wall. The ME susceptibility peaks at 266 K where $\alpha_{\parallel}^\text{max} = 3.8$ ps/m  and disappears above $T_\mathrm{N} = 308.5$ K.~\cite{borisov2007superconducting}

The source of perturbation, which is the ME energy term \eqref{eq:MEenergy}, is of the same form as the interaction energy between magnetization and an applied magnetic field, called the Zeeman energy. For electric control of the domain state, $\alpha_\parallel E \gg \mu_0 \chi_\parallel H$ must be satisfied, where $\mu_0$ is the vacuum permeability and $\chi_\parallel \approx 1.5\times10^{-3}$ is the volume magnetic susceptibility parallel to the $z$-axis close to room temperature.~\cite{foner1963high} Under this assumption, ME switching of domain in antiferromagnets is subject to reversal mechanisms analogous to magnetic field-induced magnetization reversal in ferromagnets.

\subsection{Mechanisms of magnetization reversal}

For ferromagnetic particles, magnetization reversal occurs via coherent rotation, when the particle size is smaller than the magnetostatic exchange diameter or the macrospin limit, which is about 10-50 nm depending on the material.~\cite{abo2013definition}  The magnetization dynamics is modeled by the Landau--Lifshitz--Gilbert equation,~\cite{Landau1935dispersion, gilbert2004phenomenological} which describes phenomenological damping of gyromagnetic precessions. When the particle size exceeds the exchange diameter, the magnetostatic interaction causes the spins to undergo incoherent modes of rotation such as curling and buckling for a prolate spheroid shape and fanning for a chain of spheres.~\cite{aharoni2000introduction} 

In larger ferromagnetic samples, where subdivision into domains reduces the total energy, the reversal process decomposes into local nucleation and domain wall propagation events. Reversed domains tend to nucleate at centers where the energy barrier for spin rotation may be lower, like lattice defects.~\cite{cullity2011introduction} Once the nucleation center is stabilized, the domain wall surrounding the center can propagate across the sample to spread the reversal.~\cite{vogel2006nucleation} However, the wall motion may get interrupted by the same defects, resulting in viscous damping or even a complete halt (pinning).~\cite{ferre2002} In imperfect crystals, nucleation centers may always be present, especially near edges, so that magnetization does not fully saturate. Magnetization reversal then becomes a problem of not nucleating new centers, but depinning domain walls that already exist.~\cite{cullity2011introduction} Nucleation and propagation are phenomenologically treated as thermally activated processes in Fatuzzo's theory for switching transient in ferroelectrics,~\cite{fatuzzo1962theoretical} which is later extended to magnetic films by Labrune et al.~\cite{labrune1989time} The Fatuzzo--Labrune model neglects magnetostatic effects, which is a valid assumption for antiferromagnets.

\section{Theory}
\label{Sec:theory}

Multidomain behavior in antiferromagnets can originate from domain wall pinning, without any magnetostatic interaction. For an antiferromagnetic thin-film, it is sound to consider all reversal mechanisms with non-magnetostatic origin, if the in-plane dimensions of the film are larger than the domain wall width. 

Each reversal mechanism has a coercive pressure $\mathcal{F}_\text{c}$ associated with it, which is overcome by applying ME pressure $\mathcal{F}$. If $\mathcal{F} < \mathcal{F}_\text{c}$, the mechanism is thermally activated from the metastable state with an escape time~\cite{hanggi1990reaction}
\begin{equation}
\tau = \Delta t \exp(\frac{\Delta F}{kT}),\quad \text{if}\ \Delta F \gg kT, 
\end{equation}
where $\Delta t$ is a prefactor, $\Delta F$ is the energy barrier and $kT$ is the thermal energy. For $\mathcal{F} > \mathcal{F}_\text{c}$, the energy barrier vanishes and the dynamics is governed by viscous damping in the system. From an observational perspective, the critical ME pressure for domain switching corresponds to the smallest coercive pressure among the mechanisms, if the measurement is performed slowly. The switching speed of a mechanism is distinguished from pulse response, if the pulse power overcomes the coercive pressure needed to trigger the mechanism. If the pulse activates multiple mechanisms in parallel, the fastest mechanism determines the switching speed.
\\ \\
\noindent \emph{List of symbols for magnetic properties}
\begin{symb}[noitemsep]
\item[$\mathcal{K}$]						uniaxial anisotropy of two sublattices (J/m\textsuperscript{3})
\item[$\mathcal{J}$]						antiferromagnetic exchange energy (J/m\textsuperscript{3})
\item[$\mathcal{E}$]						domain wall energy (J/m\textsuperscript{2})
\item[$M_\text{s}$]							sublattice magnetization (A/m)
\item[$\lambda$]							characteristic width of domain wall (m)
\item[$\eta$]								Gilbert damping constant
\end{symb}

\subsection{Nucleation}
Consider a finite droplet of uniformly reversed order with a domain wall boundary in an otherwise single-domain state.~\cite{vogel2006nucleation} The free energy of a cylindrical droplet of radius $r$ in a film of thickness $d$ consists of the wall energy of the curved surface and the ME energy of the reversed volume, which is written as
\begin{equation}
F_\text{N}(r) =  2\pi r d \mathcal{E}- \pi d r^2 \mathcal{F}.
\end{equation}
The energy is maximum when $\dv*{F_\text{N}}{r} = 0$, which gives $r_\text{max} = \mathcal{E}/\mathcal{F}$. If $r < r_\text{max}$ ($F_\text{N}(r)<0$), the droplet collapses because of wall pressure. If $r > r_\text{max}$ ($F_\text{N}(r)>0$) the droplet's energy decreases as it expands, so that domain reversal is set in motion. The energy barrier between $r=0$ and $r=r_\text{max}$ that needs to be surmounted to nucleate the critical radius droplet is obtained as
\begin{equation}
\Delta F_\text{N} = \frac{\pi d\mathcal{E}^2}{\mathcal{F}}.
\end{equation}
From field theory of antiferromagnets,~\cite{kim2014propulsion}  a domain wall possesses rest mass with areal density~\cite{belashchenko2016magnetoelectric}
\begin{equation}
\sigma=\frac{M_\text{s}^2}{\gamma^2\lambda\mathcal{J}},
\end{equation}
where $\gamma$ is the electron gyromagnetic ratio. The translational motion of the wall has average kinetic energy $kT/2$ from equipartition theorem, so the droplet wall fluctuates with thermal velocity
\begin{equation}
v_\text{T}(r) = \dot{r} = \sqrt{\frac{kT}{2\pi r d \sigma}}.
\label{eq:thvel}
\end{equation} 
Separate variables and integrate with the limits that correspond to $r=0$ and $r=r_\text{max}$ to find the attempt period. On substitution, the timescale of droplet nucleation is formulated as
\begin{equation}
\tau_\text{N} =  \frac{2}{3}\sqrt{\frac{2\pi d \sigma\mathcal{E}^3}{kT\mathcal{F}^3}} \exp(\frac{\pi d \mathcal{E}^2}{kT\mathcal{F}}).
\end{equation}
This equation supports the idea that nucleation is likely to occur at centers where the wall energy is locally lowered, such as dislocation lines.~\cite{aharoni1962theoretical}

\subsection{Domain wall propagation}
When a domain wall intersects a non-magnetic inclusion, the wall energy is reduced by an amount proportional to the cross-sectional area of the inclusion. Suppose a spherical inclusion with radius $R$. The interaction of antiferromagnetic wall with inclusion can be modeled in the same way as for ferromagnets when $R\gg\lambda$,~\cite{krishnan2016fundamentals} but without needing the constraint on $R$ which trivializes magnetostatic energy contribution. At a general position $r \leq R$ from the center of the inclusion, the wall energy is given as
\begin{equation}
{F}_\text{wall}(r) =  \mathcal{E}\big[S - \pi(R^2-r^2)\big],
\label{eq:Fwall}
\end{equation}
where $S\propto d$ is the natural surface area of the wall. For a relatively pure crystal, $S \gg \pi R^2$.
On applying ME pressure, the energy incurred to move the wall to a distance $r$ is given as
\begin{equation}
{F}_\text{ME}(r) = -\mathcal{F} S r.
\label{eq:FME}
\end{equation}
The positive sign of $r$ is chosen along the direction of ME stress on the wall. The total free energy, ignoring the constant terms, is reduced to
\begin{equation}
F_\text{P}(r) = \pi\mathcal{E}r^2 - \mathcal{F}Sr.
\end{equation}
In equilibrium, $\dv*{F_\text{P}}{r} = 0$ occurs at $r_\text{min} = \mathcal{F}S/(2\pi\mathcal{E})$ if $r_\text{min}<R$. The threshold ME pressure $\mathcal{F}_\text{D}$, which depins the wall away from the inclusion, is obtained by setting $r_\text{min} = R$, such that
\begin{equation}
R= \frac{\mathcal{F}_\text{D}S}{2\pi\mathcal{E}},\quad \text{if}\quad \frac{\pi R^2}{S} = \frac{\mathcal{F}_\text{D}^2S}{4\pi\mathcal{E}^2} \ll 1.
\label{eq:R}
\end{equation}
For a uniform distribution of inclusions, the analysis differs only in the calculation of cross-sectional area, which needs to be multiplied by the number of inclusions intersected by the wall. Otherwise, the distribution of inclusions in the material should  be known or accordingly, the distribution of pinning energy barriers.~\cite{ferre2002}

For $\mathcal{F} < \mathcal{F}_\text{D}$, the domain wall can thermally ``creep"~\cite{chauve2000creep} from the energy minimum at $r_\text{min}$ to the edge of the inclusion at $r = R$, by overcoming the pinning barrier
\begin{equation}
\Delta F_\text{P} = \pi \mathcal{E} R^2 \left(1 - \frac{\mathcal{F}}{\mathcal{F}_\text{D}} \right)^2 = \frac{S^2}{4\pi\mathcal{E}}(\mathcal{F}_\text{D} - \mathcal{F})^2.
\label{eq:DeltaFP}
\end{equation}
The wall moves with uniform thermal velocity 
\begin{equation}
v_\text{T} = \sqrt{\frac{kT}{\sigma S}},
\end{equation}
to give the attempt period $(R - r_\text{min})/v_\text{T}$. On substitution, the final expression for the timescale in the creep regime is found as
\begin{equation}
\tau_\text{crp} = \sqrt{\frac{\sigma S^3}{kT}}\left(\frac{\mathcal{F}_\text{D} - \mathcal{F}}{2\pi\mathcal{E}}\right) \exp[\frac{S^2(\mathcal{F}_\text{D} - \mathcal{F})^2}{4\pi kT \mathcal{E}}].
\end{equation}

For $\mathcal{F} > \mathcal{F}_\text{D}$, the domain wall motion is subject to viscous ``flow"~\cite{chauve2000creep} with linear velocity~\cite{belashchenko2016magnetoelectric}
\begin{equation}
v_\text{flw} = v_\text{T} + \frac{\eta\gamma\lambda}{\eta^2+\zeta^2}  \left(\frac{\mathcal{F} - \mathcal{F}_\text{D}}{M_\text{s}}\right),\ \ \zeta=\frac{\alpha_\parallel {E}}{\mu_0 M_\text{s}},
\label{eq:DWP}
\end{equation}
If multiple nucleation centers in the sample are activated, a wall can flow only up to a mean free path $\ell$, before it coalesces with other wall flow processes ($S\ell$ is representative of propagation activation volume in the Fatuzzo--Labrune model). The timescale in the flow regime is simply $\tau_\text{flw} =\ell/v_\text{flw}$. Since nucleation centers tend to originate around defects, the information about $\ell$ is acquired from the distribution of defects.~\cite{labrune1989time}

\subsection{Coherent rotation}
The order parameter of antiferromagnets is represented by half the difference of oppositely aligned sublattice magnetization vectors.~\cite{baltz2018antiferromagnetic} If the exchange energy is much larger than the anisotropy, the sublattice vectors would remain  antiparallel even during the reversal, so that the magnitude of the order parameter is preserved dynamically. Let  $\theta$ uniformly represent the direction of the order parameter. For a ME stress applied opposite to the initial orientation along $+z$, the free energy is described using the Stoner--Wohlfarth model~\cite{tannous2008stoner}
\begin{equation}
\frac{F_\text{SW}(\theta)}{V} = \mathcal{K}\sin^2\theta + \mu_0M_\text{s}\left(\frac{\mathcal{F}}{2\mu_0M_\text{s}}\right)\cos\theta,
\end{equation}
where $V$ is the sample volume and $\mathcal{F}/(2\mu_0M_\text{s})$ represents the staggered field experienced by sublattice magnetization. The stationary points of $F_\text{SW}$ are obtained by setting $\dv*{F_\text{SW}}{\theta} = 0$. 

To reverse when $\mathcal{F}/2<2\mathcal{K}$, the order parameter must overcome the energy barrier between the minimum at $\theta=0$ and the maximum at $\theta = \arccos(\mathcal{F}/4\mathcal{K})$ , which is calculated as
\begin{equation}
\Delta F_\text{SW} = V\mathcal{K}\left(1 - \frac{\mathcal{F}}{4\mathcal{K}}\right)^2.
\end{equation}
While the superparamagnetic limit of ferromagnetic particles is precisely formulated in the N\'eel--Brown theory,~\cite{brown1963thermal} the timescale of this limit for antiferromagnets is only heuristically proposed as~\cite{atxitia2018superparamagnetic}
\begin{equation}
\tau_\text{spm} = \frac{\eta^2+(\mathcal{K}/\mathcal{J})^2}{2\gamma\eta\mathcal{K}/M_\text{s}}\sqrt{\frac{\pi kT}{\Delta F_\text{SW}}} \exp\left(\frac{\Delta F_\text{SW}}{kT}\right).
\end{equation}
For small damping, the attempt period is diminished by the factor $(\mathcal{K/J})^2$ compared to ferromagnets, indicating faster dynamics in antiferromagnets. 

For $\mathcal{F}/2 > 2\mathcal{K}$, the reversal occurs by damping of gyromagnetic precessions. The switching time from an initial angle $\vartheta$ to final angle $\pi-\vartheta$ is given by~\cite{mallinson2000damped}
\begin{align}
\tau_\text{dmp} = \frac{\eta^2+(\mathcal{K}/\mathcal{J})^2}{\gamma\eta}& \frac{M_\text{s}}{(\mathcal{F}/2)^2 - (2\mathcal{K})^2}\left[\frac{\mathcal{F}}{2} \ln(\cot\frac{\vartheta}{2}) + \right. \nonumber \\
& \left. 2\mathcal{K} \ln(\frac{\mathcal{F}/2 - 2\mathcal{K}\cos\vartheta}{\mathcal{F}/2 + 2\mathcal{K}\cos\vartheta}) \right]. 
\end{align}
The factor $(\mathcal{K}/\mathcal{J})^2$ replaces unity in the original expression to account for antiferromagnetic dynamics.

Notice that the expressions of $\tau_\text{spm}$ and $\tau_\text{dpm}$ are derived assuming an angle-invariant $\eta$. While the Gilbert damping phenomenology explains the linewidth of magnetic resonance,~\cite{gilbert2004phenomenological} a near-equilibrium excitation, the validity of such scheme is questionable for a $180^\circ$ reversal. The phenomenology of superparamagnetism as a stochastic process called the jump-noise, captures the angular dependence of damping, including a correction to the gyromagnetic ratio.~\cite{ap2018reversal1} The jump-noise representation may also be useful to calculate the timescale in the limit of small energy barrier,~\cite{ap2018reversal2} where $\tau_\text{spm}$ diverges otherwise. Despite the merits of the jump-noise, the classical approach is still favored due to ease of extracting $\eta$ and availability of closed-form solutions.

Also, notice that $\tau_\text{dmp}$ scales indefinitely as $1/\mathcal{F}$ for large $\mathcal{F}$, which is not realistic. The ultimate limit of the switching speed is a fundamental issue concerning conservation and transfer of angular momentum and energy between the heat reservoirs of the system: spins and lattice.~\cite{stohr2007magnetism} The exploration of ultrafast dynamics in antiferromagnets, however, is still in its infancy.~\cite{kirilyuk2010ultrafast}

\section{Application}
\label{Sec:appl}

\begin{figure}
\includegraphics[scale=1]{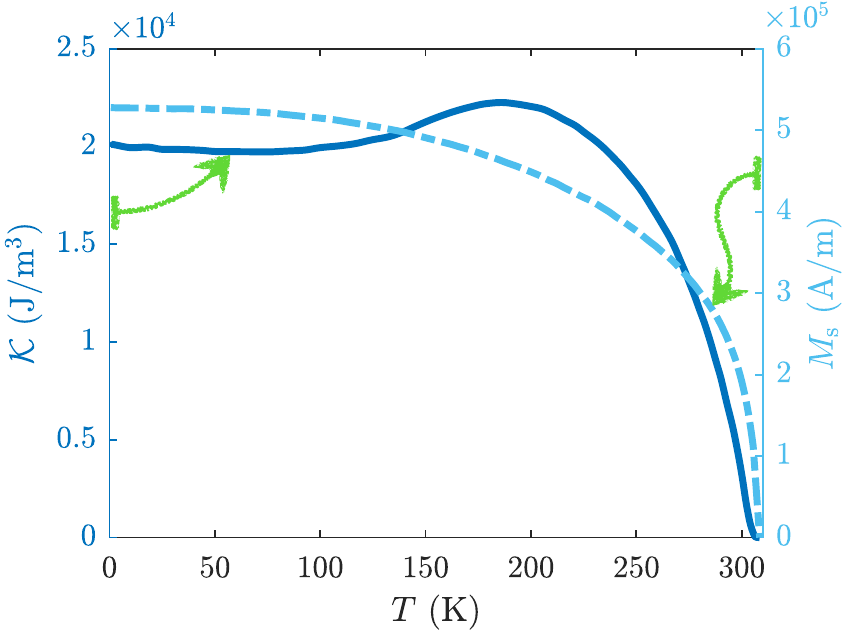}
\caption{Uniaxial anisotropy and sublattice magnetization of Cr\textsubscript{2}O\textsubscript{3} vs temperature remastered from Refs.~\onlinecite{foner1963high}~and~\onlinecite{samuelsen1970inelastic}, respectively. The y-axis is color-mapped to the plot.}
\label{fig:KMs}
\end{figure}

To assess ME reversal of antiferromagnetic domain, a number sense of the energy barriers and timescales involved with reversal mechanisms is necessary. Consider the two recent works~\cite{kosub2017purely, toyoki2015magnetoelectric} on ME switching of thin-film Cr\textsubscript{2}O\textsubscript{3}, which have identical film thickness. The specifics of the experiments are summarized in Table~\ref{tab:expt}. The switching in Ref.~\onlinecite{toyoki2015magnetoelectric} is observed for the case where the exchange bias aids in domain reversal. So, the net external pressure is the sum of applied ME and exchange bias pressures. The coercive ME pressure in Ref.~\onlinecite{kosub2017purely} is significantly lower because the temperature is much closer to $T_\text{N}$. The material parameters of bulk Cr\textsubscript{2}O\textsubscript{3} at relevant temperatures are listed in Table~\ref{tab:Physprop}. The values of $\mathcal{K}$ and $M_\text{s}$ are extracted from Fig.~\ref{fig:KMs}.

\begin{table}
\caption{\label{tab:expt} Experiments on ME switching of thin-film Cr\textsubscript{2}O\textsubscript{3}}
\begin{ruledtabular}
\begin{tabular}{@{} l l l @{}}
		& 		\textbf{Ref.~\onlinecite{toyoki2015magnetoelectric}}			&		\textbf{Ref.~\onlinecite{kosub2017purely}}\\[1pt]
\hline\\[-7pt]
Type		&		exchange bias			&		purely antiferromagnet\\[1pt]
$d$		&		200 nm						&		200 nm\\[1pt]
$V$		&		---							&		$d \times$(0.1 cm\textsuperscript{2})	\\[1pt]
$T$		&		268 K						&		292 K\\[1pt]
$H$\textsuperscript{$\ast$}	&		0.798 MA/m				&		0.5 MA/m\\[1pt]	
$E_\text{c}$\textsuperscript{$\ast\ast$}		&		125 V/\textmu m		&		7.50 V/\textmu m\\[1pt]				
$\alpha_\parallel$~\cite{borisov2007superconducting}		&		$3.7$ ps/m		&		$3.3$ ps/m\\[1pt]
$\mathcal{F}_\text{EB}$\textsuperscript{$\star$}		&		930 J/m\textsuperscript{3}		& 		$\varnothing$\\[1pt]
$\mathcal{F}_\text{c}$\textsuperscript{$\dagger$}		&			1.67 kJ/m\textsuperscript{3}		&		24.8 J/m\textsuperscript{3}\\[1pt]
$\tau$	&		0.1 \textmu s				&		---\\[1pt]							
\end{tabular}
\end{ruledtabular}
\vspace{-10pt}
\begin{flushright}
\textsuperscript{$\ast$}constant magnetic field\hfill
\textsuperscript{$\ast\ast$}coercive electric field\\
\textsuperscript{$\star$}exchange bias pressure\hfill
\textsuperscript{$\dagger$}$\mathcal{F}_\text{c} = 2\alpha_\parallel E_\text{c} H + \mathcal{F}_\text{EB}$
\end{flushright}
\end{table}

\begin{table}
 \caption{\label{tab:Physprop} Material properties of bulk Cr\textsubscript{2}O\textsubscript{3}}
 \begin{ruledtabular}
 \begin{tabular}{@{} l l l l @{}}
\textbf{Symbol} & $\boldsymbol{T= 4.2}$ \textbf{K} & $\boldsymbol{T=268}$ \textbf{K} & $\boldsymbol{T=292}$ \textbf{K} \\[1pt]
\hline\\[-7pt]
$\mathcal{K}$~\cite{foner1963high} 		&  		$20$ kJ/m\textsuperscript{3}			&			$15$ kJ/m\textsuperscript{3}		&		$7.3$ kJ/m\textsuperscript{3}\\[1pt]
$M_\text{s}$~\cite{samuelsen1970inelastic, artman1965magnetic}			&		$0.58$ MA/m 		&		$0.36$ MA/m		&		$0.26$ MA/m\\[1pt]\cline{3-4}\\[-9pt]
$\mathcal{J}$~\cite{foner1963high}				&		$280$ MJ/m\textsuperscript{3}		&			\multicolumn{2}{c}{$\propto M_\text{s}$}\\[1pt]
$\mathcal{E}$~\cite{belashchenko2016magnetoelectric}			&			$0.96$ mJ/m\textsuperscript{2}		&		 \multicolumn{2}{c}{$\propto \sqrt{\mathcal{K}}$}\\[1pt]
$\lambda$~\cite{belashchenko2016magnetoelectric}		&		$12$ nm		& 		\multicolumn{2}{c}{$\propto 1/\sqrt{\mathcal{K}}$}\\[1pt]\cline{2-4}\\[-8pt]
$\eta$~\cite{wang2015spin}		&		\multicolumn{3}{c}{$1.2\times 10^{-3}$}\\[1pt]
 \end{tabular}
 \end{ruledtabular}
 \end{table}

\begin{figure}
\includegraphics[scale=1]{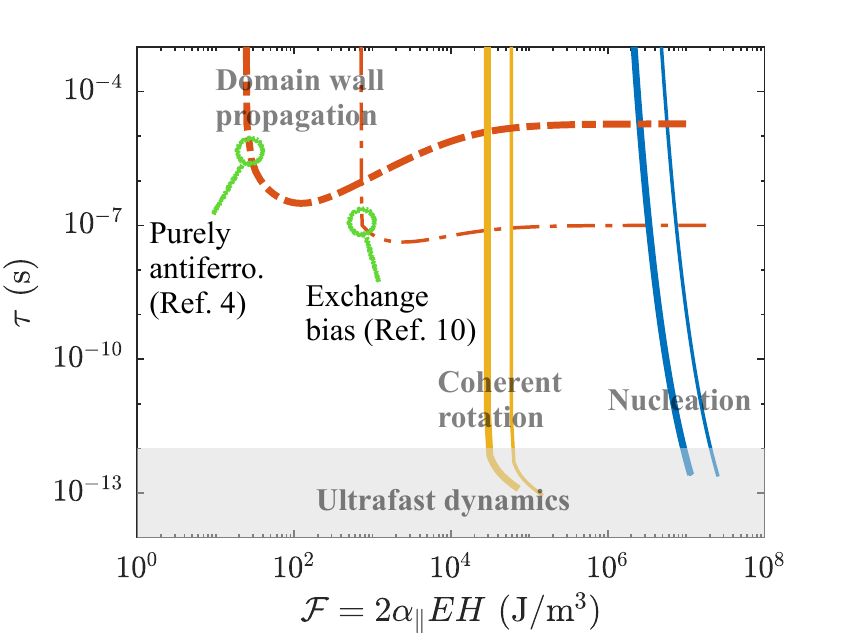}
\caption{Timescales of domain reversal mechanisms versus applied ME pressure for thin-film Cr\textsubscript{2}O\textsubscript{3} in constant magnetic field. The thinner and thicker plot lines respectively correspond to the experiments of Refs.~\onlinecite{toyoki2015magnetoelectric}~and~\onlinecite{kosub2017purely} (Table~\ref{tab:expt}), evaluated using material properties in Table.~\ref{tab:Physprop} and domain wall propagation parameters in Table~\ref{tab:DWpara}. The gray region represents the timescale of ultrafast spin and lattice dynamics.~\cite{satoh2007ultrafast} The small circle marks the operating point of the experiments. Dielectric breakdown in thin-film Cr\textsubscript{2}O\textsubscript{3} occcurs at an electric field of 200 V/\textmu m.~\cite{sun2017local}}
\label{fig:Ftau}
\end{figure}

Using the information in Tables~\ref{tab:expt}~and~\ref{tab:Physprop}, the timescales of nucleation and coherent rotation (assume $\vartheta=1^\circ$ and same $V$ for Ref.~\onlinecite{toyoki2015magnetoelectric}) can be evaluated as shown in Fig.~\ref{fig:Ftau}. The energy barrier of both mechanisms is orders of magnitude larger than the measured coercive ME pressure, so they are suppressed. An earlier work~\cite{brown1969domain} indeed observed that a very large ME pressure tends to produce a fully saturated state that could not be switched further. Therefore, at moderate ME pressures, domain wall propagation is the responsible mechanism for reversal. 

Estimating the timescale of domain wall propagation requires additional knowledge of the wall surface area $S$ and mean free path $\ell$. Because of symmetry of wall direction in the film plane, it is hypothesized that $S= \ell d$.  At the depinning  pressure $\mathcal{F}_\text{d} = \mathcal{F}_\text{c}$, $\ell = v_\text{T}\tau$ so that $S= d v_\text{T}\tau$. Using Eq.~\eqref{eq:thvel}, 
\begin{equation}
S = \left(\frac{kTd^2\tau^2}{\sigma}\right)^{1/3}.
\end{equation}
Since $\tau$ is known for the experiment of Ref.~\onlinecite{toyoki2015magnetoelectric} (Table~\ref{tab:expt}), $\ell$, $S$ and the inclusion radius $R$ \eqref{eq:R} can be calculated. If the same $R$ is assumed for the sample of Ref.~\onlinecite{kosub2017purely}, $\ell$ and $S$ can be estimated by retracing the steps. The values of $\ell$, $R$ and error \eqref{eq:R} are listed in Table~\ref{tab:DWpara}. Note that $\pi R^2$ represents the total cross-sectional area of the inclusions that intersect the wall, not that of a single inclusion, so $R>d/2$ is allowed as long as $\pi R^2 < S$ . Using the parameters in Table~\ref{tab:DWpara}, the timescale of wall propagation is plotted in Fig.~\ref{fig:Ftau} (the non-monotonic behavior agrees with Eq.~\ref{eq:DWP}). The switching speed for Ref.~\onlinecite{kosub2017purely} is estimated as 10 \textmu s, which is different from the stated $\sim 10$ ns without measurement or theoretical validation. 

\begin{table}
\caption{\label{tab:DWpara} Estimated domain wall propagation parameters for experiments on ME switching of thin-film Cr\textsubscript{2}O\textsubscript{3}}
\begin{ruledtabular}
\begin{tabular}{@{} l l l @{}}
Parameters		& 		\textbf{Ref.~\onlinecite{toyoki2015magnetoelectric}}			&		\textbf{Ref.~\onlinecite{kosub2017purely}}\\[1pt]
\hline\\[-7pt]
$\ell$		&		$4.8$ \textmu m		&		220 \textmu m\\[1pt]\cline{2-3}\\[-9pt]
$R$		&		\multicolumn{2}{c}{$300$ nm}\\[1pt]\cline{2-3}\\[-9pt]
$\pi R^2/S$		&		0.31			&		$6.6\times 10^{-3}$
\end{tabular}
\end{ruledtabular}
\end{table}

Below picosecond timescale, the spin-lattice interaction in Cr\textsubscript{2}O\textsubscript{3} comes into play,~\cite{satoh2007ultrafast} and the models for nucleation and coherent rotation would become invalid. In addition to the breakdown of the models, the electric fields required to achieve ultrafast dynamics is likely to induce dielectric breakdown. In thin-films of Cr\textsubscript{2}O\textsubscript{3}, the breakdown electric field drops from 1000 V/\textmu m of bulk to 200 V/\textmu m.~\cite{sun2017local} Even at a generous magnetic field of 1 T, the ME pressure that can be reached before breakdown is just 525 J/m\textsuperscript{3}. Experimental advancements in increasing the breakdown field and reducing the value of the magnetic anisotropy~\cite{mu2018influence} will be needed to realize dynamics in the sub-nanosecond regime.

\section{Conclusion}

In this paper, we theoretically analyze domain reversal in thin-film ME antiferromagnets in the form of nucleation, domain wall propagation and coherent rotation. We model the timescales of the reversal mechanisms as a function of applied ME pressure, with the knowledge of material properties and domain wall propagation parameters. We apply the theory to recent experimental works~\cite{toyoki2015magnetoelectric, kosub2017purely} on ME switching of thin-film Cr\textsubscript{2}O\textsubscript{3}, which have identical film-thickness. We find that nucleation and coherent rotation are suppressed at the ME pressures used in the experiments, and domain wall propagation is responsible for reversal.\footnote{At the time of submission of manuscript, we came across experimental works published  less than a month before by Shiratsuchi et al.,~\cite{shiratsuchi2018observation, shiratsuchi2018antiferromagnetic, nguyen2018energy}  which corroborate the role of domain wall propagation in ME switching of exchange bias. We will assess our theory with their findings in a future work.} We extract wall propagation parameters from the experiment of Ref.~\onlinecite{toyoki2015magnetoelectric} and estimate a switching speed of 10 \textmu s for Ref.~\onlinecite{kosub2017purely}, which has not been measured. We remark that ultrafast dynamics of ME antiferromagnets is limited by low dielectric breakdown and high magnetic anisotropy. Experiments should focus on mitigating these limitations in order to unleash switching at terahertz speeds.

\appendix
\section{Timeline of ME research in \texorpdfstring{Cr\textsubscript{2}O\textsubscript{3}}{Cr2O3}}
\label{Sec:timeline}
The ME effect in Cr\textsubscript{2}O\textsubscript{3} was predicted phenomenologically by Dzyaloshinskii in 1959,~\cite{dzyaloshinskii1960magneto} which was soon followed by measurement of the temperature-dependent ME susceptibility.~\cite{astrov1961magnetoelectric, folen1961anisotropy}  A few years later, the domain switching in Cr\textsubscript{2}O\textsubscript{3} was demonstrated by simultaneous application of electric and magnetic fields to a bulk sample.~\cite{martin1966antiferromagnetic, brown1969domain} The direction of domain was detected directly by monitoring the ME susceptibility during the switching process; the domain state could not be read in the absence of applied fields. The speed of domain reversal was measured similar to 1 ms for an applied ME pressure of about 2 J/m\textsuperscript{3} close to room temperature. With an electric-ﬁeld of $10^6$ V/cm, the magnetization obtained from ME effect corresponds to reversal of only five of every one million spins! From the 1970s, the research in ME started declining because the direct coupling was considered too weak for ME phase control in storage elements.~\cite{fiebig2005revival}

Fast-forward nearly four decades, the interest in ME research revived. The ME switching of exchange bias in a ferromagnetic layer interfacially, coupled to Cr\textsubscript{2}O\textsubscript{3}  bulk, was achieved after ME field cooling through the N\'eel temperature.~\cite{borisov2005magnetoelectric} The one-to-one coupling between ferromagnet’s magnetization and antiferromagnetic order made it possible to readout the antiferromagnetic domain state independently from the switching process. Five years later, the same was accomplished isothermally at room temperature.~\cite{he2010robust} Equilibrium magnetization at the boundary of ME antiferromagnet was proven from symmetry arguments,~\cite{belashchenko2010equilibrium} and also imaged on the surface of Cr\textsubscript{2}O\textsubscript{3} film.~\cite{wu2011imaging} There were theoretical~\cite{mu2013effect} and practical~\cite{street2014increasing} works foussing on enhancing the N\'eel temperature. Research efforts from a different group confirmed isothermal switching of the exchange bias as well as the magnetization in zero magnetic field in an all-thin-film system.~\cite{ashida2015isothermal} In a parallel work,~\cite{toyoki2015magnetoelectric} the switching speed was additionally investigated by pulse voltage measurements, and found to be around 0.1 \textmu s. In the recent study,~\cite{kosub2017purely} an all-electric access to the surface magnetization was demonstrated via anomalous Hall magnetometry~\cite{kosub2015all} for ferromagnet-free, thin-film ME  memory. The switching speed was not examined, but prospected to be within a few tens of nanosecond. 

Theoretical works have made attempts to explain the microscopic origin of the ME effect in equilibrium in Cr\textsubscript{2}O\textsubscript{3}. The earlier works~\cite{rado1961mechanism, date1961origin, hornreich1967statistical} were mostly phenomenolgical, while the studies during the last decade~\cite{mostovoy2010temperature, bousquet2011unexpectedly, malashevich2012full} were based on first-principles calculations. However, there is no clear consensus on the cause of ME effect yet.

\section{Issues with domain formation in \texorpdfstring{Cr\textsubscript{2}O\textsubscript{3}}{Cr2O3}}
\label{Sec:issues}

\begin{figure}
\centering
\includegraphics[width=2in]{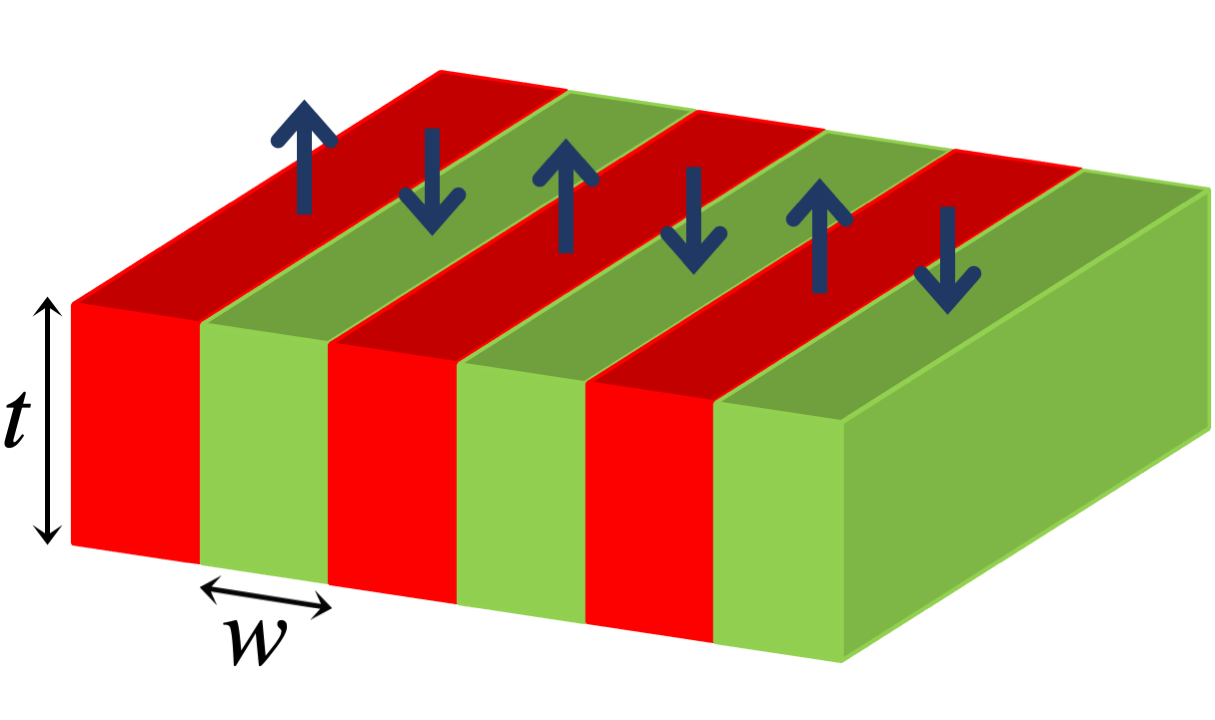}
\caption{Conceptual domain structure of Cr\textsubscript{2}O\textsubscript{3} film, where the top surface corresponds to the (0001) plane. The red and green colors represent the two antiferromagnetic orders, respectively. Arrows on top of each stripe indicate the direction of the surface magnetization coupled to the bulk order.}
\label{fig:CPS}
\end{figure}

Consider the thin-film sample of Ref.~\onlinecite{wu2011imaging}, which is 127 nm thick, and shows stripe-like surface magnetization domains, that are 1 {\textmu m} wide, on the (0001) plane (see Fig.~\ref{fig:CPS} for an illustration).  The image of the boundary uncovers the domain structure inside the bulk because the boundary magnetization is coupled to the bulk order in a one-to-one manner.~\cite{andreev1996macroscopic, belashchenko2010equilibrium} 

Suppose that the non-trivial domain structure is formed because of the significance of the magnetostatic energy due to surface magnetization. Cr\textsubscript{2}O\textsubscript{3} has $R\overline{3}c$ space group with lattice constant $a=0.5$ nm.~\cite{Villars2016} There are six magnetic Cr\textsuperscript{3+} ions in the unit cell in the (0001) plane, with spins aligned along the trigonal $z$-axis.  Cr\textsuperscript{3+} has the electronic configuration 3d\textsuperscript{3}, so each ion can at most contribute to 3$\mu_\text{B}$ magnetic moment. Therefore, the areal magnetic moment of the film is calculated as
\begin{equation}
\mathcal{M} = \frac{6\times 3\mu_\text{B}}{\sqrt{3}/2 a^2} = 7.83\times 10^{-4}\ \text{A}.
\label{eq:Ms}
\end{equation}
For a film of thickness $d$, the magnetostatic energy $\mathcal{E}_\text{ms}$ averaged over the sample volume is then obtained as
\begin{equation}
\mathcal{F}_\text{ms} = \frac{1}{2}\mu_0 \left(\frac{\mathcal{M}}{d}\right)^2 = (3.85\times 10^{-13})\cdot\frac{1}{d^2}\ \text{J/m\textsuperscript{3}}
\end{equation}
Multidomain structures in antiferromagnets, once formed, follow the same constraints of exchange and anisotropy as their ferromagnetic counterparts and possess domain wall energy, which is calculated as~\cite{foner1963high, belashchenko2016magnetoelectric} 
\begin{equation}
\mathcal{E} =  4\mathcal{K}\lambda = 9.6 \times 10^{-4}\ \text{J/m\textsuperscript{2}},
\label{eq:gammawall}
\end{equation}
where $\mathcal{K} = 2\times 10^4 $ J/m\textsuperscript{3} is the combined uniaxial anisotropy of two sublattices and $\lambda=\sqrt{\mathcal{A/K}} = 12$ nm is the characteristic wall length. For an average domain width $w$, the domain wall energy density per unit volume is $\mathcal{F}_\text{wall} = \mathcal{E}/w$.

For the parameters $d = 127$ nm and $w = 1$ {\textmu m}, $\mathcal{F}_\text{ms} = 23.9$ \text{J/m\textsuperscript{3}}, whereas $\mathcal{F}_\text{wall} = 960$ \text{J/m\textsuperscript{3}}. Wall formation overall costs a lot more energy because the magnetostatic contribution is negligible. For a defect-free sample, multidomain structures in antiferromagnets are not expected.

\begin{acknowledgments}
This work was supported in part by the Semiconductor Research Corporation (SRC) and the National Science Foundation (NSF) through ECCS 1740136. S. Rakheja also acknowledges the funding support from the MRSEC Program of the National Science Foundation under Award Number DMR-1420073.
\end{acknowledgments}

\bibliography{refs}

\providecommand{\noopsort}[1]{}\providecommand{\singleletter}[1]{#1}%
\begin{thebibliography}{67}%
\makeatletter
\providecommand \@ifxundefined [1]{%
 \@ifx{#1\undefined}
}%
\providecommand \@ifnum [1]{%
 \ifnum #1\expandafter \@firstoftwo
 \else \expandafter \@secondoftwo
 \fi
}%
\providecommand \@ifx [1]{%
 \ifx #1\expandafter \@firstoftwo
 \else \expandafter \@secondoftwo
 \fi
}%
\providecommand \natexlab [1]{#1}%
\providecommand \enquote  [1]{``#1''}%
\providecommand \bibnamefont  [1]{#1}%
\providecommand \bibfnamefont [1]{#1}%
\providecommand \citenamefont [1]{#1}%
\providecommand \href@noop [0]{\@secondoftwo}%
\providecommand \href [0]{\begingroup \@sanitize@url \@href}%
\providecommand \@href[1]{\@@startlink{#1}\@@href}%
\providecommand \@@href[1]{\endgroup#1\@@endlink}%
\providecommand \@sanitize@url [0]{\catcode `\\12\catcode `\$12\catcode
  `\&12\catcode `\#12\catcode `\^12\catcode `\_12\catcode `\%12\relax}%
\providecommand \@@startlink[1]{}%
\providecommand \@@endlink[0]{}%
\providecommand \url  [0]{\begingroup\@sanitize@url \@url }%
\providecommand \@url [1]{\endgroup\@href {#1}{\urlprefix }}%
\providecommand \urlprefix  [0]{URL }%
\providecommand \Eprint [0]{\href }%
\providecommand \doibase [0]{http://dx.doi.org/}%
\providecommand \selectlanguage [0]{\@gobble}%
\providecommand \bibinfo  [0]{\@secondoftwo}%
\providecommand \bibfield  [0]{\@secondoftwo}%
\providecommand \translation [1]{[#1]}%
\providecommand \BibitemOpen [0]{}%
\providecommand \bibitemStop [0]{}%
\providecommand \bibitemNoStop [0]{.\EOS\space}%
\providecommand \EOS [0]{\spacefactor3000\relax}%
\providecommand \BibitemShut  [1]{\csname bibitem#1\endcsname}%
\let\auto@bib@innerbib\@empty
\bibitem [{\citenamefont {Fiebig}(2005)}]{fiebig2005revival}%
  \BibitemOpen
  \bibfield  {author} {\bibinfo {author} {\bibfnamefont {M.}~\bibnamefont
  {Fiebig}},\ }\href@noop {} {\bibfield  {journal} {\bibinfo  {journal}
  {Journal of Physics D: Applied Physics}\ }\textbf {\bibinfo {volume} {38}},\
  \bibinfo {pages} {R123} (\bibinfo {year} {2005})}\BibitemShut {NoStop}%
\bibitem [{\citenamefont {Manipatruni}\ \emph {et~al.}(2018)\citenamefont
  {Manipatruni}, \citenamefont {Nikonov}, \citenamefont {Lin}, \citenamefont
  {Gosavi}, \citenamefont {Liu}, \citenamefont {Prasad}, \citenamefont {Huang},
  \citenamefont {Bonturim}, \citenamefont {Ramesh},\ and\ \citenamefont
  {Young}}]{manipatruni2018scalable}%
  \BibitemOpen
  \bibfield  {author} {\bibinfo {author} {\bibfnamefont {S.}~\bibnamefont
  {Manipatruni}}, \bibinfo {author} {\bibfnamefont {D.~E.}\ \bibnamefont
  {Nikonov}}, \bibinfo {author} {\bibfnamefont {C.-C.}\ \bibnamefont {Lin}},
  \bibinfo {author} {\bibfnamefont {T.~A.}\ \bibnamefont {Gosavi}}, \bibinfo
  {author} {\bibfnamefont {H.}~\bibnamefont {Liu}}, \bibinfo {author}
  {\bibfnamefont {B.}~\bibnamefont {Prasad}}, \bibinfo {author} {\bibfnamefont
  {Y.-L.}\ \bibnamefont {Huang}}, \bibinfo {author} {\bibfnamefont
  {E.}~\bibnamefont {Bonturim}}, \bibinfo {author} {\bibfnamefont
  {R.}~\bibnamefont {Ramesh}}, \ and\ \bibinfo {author} {\bibfnamefont {I.~A.}\
  \bibnamefont {Young}},\ }\href@noop {} {\bibfield  {journal} {\bibinfo
  {journal} {Nature}\ ,\ \bibinfo {pages} {1}} (\bibinfo {year}
  {2018})}\BibitemShut {NoStop}%
\bibitem [{\citenamefont {Dowben}\ \emph {et~al.}(2018)\citenamefont {Dowben},
  \citenamefont {Binek}, \citenamefont {Zhang}, \citenamefont {Wang},
  \citenamefont {Mei}, \citenamefont {Bird}, \citenamefont {Singisetti},
  \citenamefont {Hong}, \citenamefont {Wang},\ and\ \citenamefont
  {Nikonov}}]{dowben2018towards}%
  \BibitemOpen
  \bibfield  {author} {\bibinfo {author} {\bibfnamefont {P.~A.}\ \bibnamefont
  {Dowben}}, \bibinfo {author} {\bibfnamefont {C.}~\bibnamefont {Binek}},
  \bibinfo {author} {\bibfnamefont {K.}~\bibnamefont {Zhang}}, \bibinfo
  {author} {\bibfnamefont {L.}~\bibnamefont {Wang}}, \bibinfo {author}
  {\bibfnamefont {W.-N.}\ \bibnamefont {Mei}}, \bibinfo {author} {\bibfnamefont
  {J.~P.}\ \bibnamefont {Bird}}, \bibinfo {author} {\bibfnamefont
  {U.}~\bibnamefont {Singisetti}}, \bibinfo {author} {\bibfnamefont
  {X.}~\bibnamefont {Hong}}, \bibinfo {author} {\bibfnamefont {K.~L.}\
  \bibnamefont {Wang}}, \ and\ \bibinfo {author} {\bibfnamefont
  {D.}~\bibnamefont {Nikonov}},\ }\href@noop {} {\bibfield  {journal} {\bibinfo
   {journal} {IEEE Journal on Exploratory Solid-State Computational Devices and
  Circuits}\ } (\bibinfo {year} {2018})}\BibitemShut {NoStop}%
\bibitem [{\citenamefont {Kosub}\ \emph {et~al.}(2017)\citenamefont {Kosub},
  \citenamefont {Kopte}, \citenamefont {H{\"u}hne}, \citenamefont {Appel},
  \citenamefont {Shields}, \citenamefont {Maletinsky}, \citenamefont
  {H{\"u}bner}, \citenamefont {Liedke}, \citenamefont {Fassbender},
  \citenamefont {Schmidt} \emph {et~al.}}]{kosub2017purely}%
  \BibitemOpen
  \bibfield  {author} {\bibinfo {author} {\bibfnamefont {T.}~\bibnamefont
  {Kosub}}, \bibinfo {author} {\bibfnamefont {M.}~\bibnamefont {Kopte}},
  \bibinfo {author} {\bibfnamefont {R.}~\bibnamefont {H{\"u}hne}}, \bibinfo
  {author} {\bibfnamefont {P.}~\bibnamefont {Appel}}, \bibinfo {author}
  {\bibfnamefont {B.}~\bibnamefont {Shields}}, \bibinfo {author} {\bibfnamefont
  {P.}~\bibnamefont {Maletinsky}}, \bibinfo {author} {\bibfnamefont
  {R.}~\bibnamefont {H{\"u}bner}}, \bibinfo {author} {\bibfnamefont {M.~O.}\
  \bibnamefont {Liedke}}, \bibinfo {author} {\bibfnamefont {J.}~\bibnamefont
  {Fassbender}}, \bibinfo {author} {\bibfnamefont {O.~G.}\ \bibnamefont
  {Schmidt}},  \emph {et~al.},\ }\href@noop {} {\bibfield  {journal} {\bibinfo
  {journal} {Nature Communications}\ }\textbf {\bibinfo {volume} {8}},\
  \bibinfo {pages} {13985} (\bibinfo {year} {2017})}\BibitemShut {NoStop}%
\bibitem [{\citenamefont {Martin}\ and\ \citenamefont
  {Anderson}(1966)}]{martin1966antiferromagnetic}%
  \BibitemOpen
  \bibfield  {author} {\bibinfo {author} {\bibfnamefont {T.}~\bibnamefont
  {Martin}}\ and\ \bibinfo {author} {\bibfnamefont {J.}~\bibnamefont
  {Anderson}},\ }\href@noop {} {\bibfield  {journal} {\bibinfo  {journal} {IEEE
  Transactions on Magnetics}\ }\textbf {\bibinfo {volume} {2}},\ \bibinfo
  {pages} {446} (\bibinfo {year} {1966})}\BibitemShut {NoStop}%
\bibitem [{\citenamefont {Li}(1956)}]{li1956domain}%
  \BibitemOpen
  \bibfield  {author} {\bibinfo {author} {\bibfnamefont {Y.-Y.}\ \bibnamefont
  {Li}},\ }\href@noop {} {\bibfield  {journal} {\bibinfo  {journal} {Physical
  Review}\ }\textbf {\bibinfo {volume} {101}},\ \bibinfo {pages} {1450}
  (\bibinfo {year} {1956})}\BibitemShut {NoStop}%
\bibitem [{\citenamefont {Fiebig}\ \emph {et~al.}(1995)\citenamefont {Fiebig},
  \citenamefont {Fr{\"o}hlich}, \citenamefont {Sluyterman~v. L},\ and\
  \citenamefont {Pisarev}}]{fiebig1995domain}%
  \BibitemOpen
  \bibfield  {author} {\bibinfo {author} {\bibfnamefont {M.}~\bibnamefont
  {Fiebig}}, \bibinfo {author} {\bibfnamefont {D.}~\bibnamefont
  {Fr{\"o}hlich}}, \bibinfo {author} {\bibfnamefont {G.}~\bibnamefont
  {Sluyterman~v. L}}, \ and\ \bibinfo {author} {\bibfnamefont {R.}~\bibnamefont
  {Pisarev}},\ }\href@noop {} {\bibfield  {journal} {\bibinfo  {journal}
  {Applied Physics Letters}\ }\textbf {\bibinfo {volume} {66}},\ \bibinfo
  {pages} {2906} (\bibinfo {year} {1995})}\BibitemShut {NoStop}%
\bibitem [{\citenamefont {Wu}\ \emph {et~al.}(2011)\citenamefont {Wu},
  \citenamefont {He}, \citenamefont {Wysocki}, \citenamefont {Lanke},
  \citenamefont {Komesu}, \citenamefont {Belashchenko}, \citenamefont {Binek},\
  and\ \citenamefont {Dowben}}]{wu2011imaging}%
  \BibitemOpen
  \bibfield  {author} {\bibinfo {author} {\bibfnamefont {N.}~\bibnamefont
  {Wu}}, \bibinfo {author} {\bibfnamefont {X.}~\bibnamefont {He}}, \bibinfo
  {author} {\bibfnamefont {A.~L.}\ \bibnamefont {Wysocki}}, \bibinfo {author}
  {\bibfnamefont {U.}~\bibnamefont {Lanke}}, \bibinfo {author} {\bibfnamefont
  {T.}~\bibnamefont {Komesu}}, \bibinfo {author} {\bibfnamefont {K.~D.}\
  \bibnamefont {Belashchenko}}, \bibinfo {author} {\bibfnamefont
  {C.}~\bibnamefont {Binek}}, \ and\ \bibinfo {author} {\bibfnamefont {P.~A.}\
  \bibnamefont {Dowben}},\ }\href@noop {} {\bibfield  {journal} {\bibinfo
  {journal} {Physical Review Letters}\ }\textbf {\bibinfo {volume} {106}},\
  \bibinfo {pages} {087202} (\bibinfo {year} {2011})}\BibitemShut {NoStop}%
\bibitem [{\citenamefont {Brown}(1969)}]{brown1969magneto}%
  \BibitemOpen
  \bibfield  {author} {\bibinfo {author} {\bibfnamefont {C.~A.}\ \bibnamefont
  {Brown}},\ }\emph {\bibinfo {title} {Magneto-electric domains in single
  crystal chromium oxide}},\ \href@noop {} {\bibinfo {type} {{Ph.D.} thesis}},\
  \bibinfo  {school} {Imperial College London} (\bibinfo {year}
  {1969})\BibitemShut {NoStop}%
\bibitem [{\citenamefont {Toyoki}\ \emph {et~al.}(2015)\citenamefont {Toyoki},
  \citenamefont {Shiratsuchi}, \citenamefont {Kobane}, \citenamefont
  {Mitsumata}, \citenamefont {Kotani}, \citenamefont {Nakamura},\ and\
  \citenamefont {Nakatani}}]{toyoki2015magnetoelectric}%
  \BibitemOpen
  \bibfield  {author} {\bibinfo {author} {\bibfnamefont {K.}~\bibnamefont
  {Toyoki}}, \bibinfo {author} {\bibfnamefont {Y.}~\bibnamefont {Shiratsuchi}},
  \bibinfo {author} {\bibfnamefont {A.}~\bibnamefont {Kobane}}, \bibinfo
  {author} {\bibfnamefont {C.}~\bibnamefont {Mitsumata}}, \bibinfo {author}
  {\bibfnamefont {Y.}~\bibnamefont {Kotani}}, \bibinfo {author} {\bibfnamefont
  {T.}~\bibnamefont {Nakamura}}, \ and\ \bibinfo {author} {\bibfnamefont
  {R.}~\bibnamefont {Nakatani}},\ }\href@noop {} {\bibfield  {journal}
  {\bibinfo  {journal} {Applied Physics Letters}\ }\textbf {\bibinfo {volume}
  {106}},\ \bibinfo {pages} {162404} (\bibinfo {year} {2015})}\BibitemShut
  {NoStop}%
\bibitem [{\citenamefont {Shtrikman}\ and\ \citenamefont
  {Treves}(1960)}]{shtrikman1960resolution}%
  \BibitemOpen
  \bibfield  {author} {\bibinfo {author} {\bibfnamefont {S.}~\bibnamefont
  {Shtrikman}}\ and\ \bibinfo {author} {\bibfnamefont {D.}~\bibnamefont
  {Treves}},\ }\href@noop {} {\bibfield  {journal} {\bibinfo  {journal}
  {Journal of Applied Physics}\ }\textbf {\bibinfo {volume} {31}},\ \bibinfo
  {pages} {S72} (\bibinfo {year} {1960})}\BibitemShut {NoStop}%
\bibitem [{\citenamefont {Ashida}\ \emph {et~al.}(2014)\citenamefont {Ashida},
  \citenamefont {Oida}, \citenamefont {Shimomura}, \citenamefont {Nozaki},
  \citenamefont {Shibata},\ and\ \citenamefont
  {Sahashi}}]{ashida2014observation}%
  \BibitemOpen
  \bibfield  {author} {\bibinfo {author} {\bibfnamefont {T.}~\bibnamefont
  {Ashida}}, \bibinfo {author} {\bibfnamefont {M.}~\bibnamefont {Oida}},
  \bibinfo {author} {\bibfnamefont {N.}~\bibnamefont {Shimomura}}, \bibinfo
  {author} {\bibfnamefont {T.}~\bibnamefont {Nozaki}}, \bibinfo {author}
  {\bibfnamefont {T.}~\bibnamefont {Shibata}}, \ and\ \bibinfo {author}
  {\bibfnamefont {M.}~\bibnamefont {Sahashi}},\ }\href@noop {} {\bibfield
  {journal} {\bibinfo  {journal} {Applied Physics Letters}\ }\textbf {\bibinfo
  {volume} {104}},\ \bibinfo {pages} {152409} (\bibinfo {year}
  {2014})}\BibitemShut {NoStop}%
\bibitem [{\citenamefont {Dzyaloshinskii}(1960)}]{dzyaloshinskii1960magneto}%
  \BibitemOpen
  \bibfield  {author} {\bibinfo {author} {\bibfnamefont {I.}~\bibnamefont
  {Dzyaloshinskii}},\ }\href@noop {} {\bibfield  {journal} {\bibinfo  {journal}
  {Soviet Physics Journal of Experimental and Theoretical Physics}\ }\textbf
  {\bibinfo {volume} {10}},\ \bibinfo {pages} {628} (\bibinfo {year}
  {1960})}\BibitemShut {NoStop}%
\bibitem [{\citenamefont {Mostovoy}\ \emph {et~al.}(2010)\citenamefont
  {Mostovoy}, \citenamefont {Scaramucci}, \citenamefont {Spaldin},\ and\
  \citenamefont {Delaney}}]{mostovoy2010temperature}%
  \BibitemOpen
  \bibfield  {author} {\bibinfo {author} {\bibfnamefont {M.}~\bibnamefont
  {Mostovoy}}, \bibinfo {author} {\bibfnamefont {A.}~\bibnamefont
  {Scaramucci}}, \bibinfo {author} {\bibfnamefont {N.~A.}\ \bibnamefont
  {Spaldin}}, \ and\ \bibinfo {author} {\bibfnamefont {K.~T.}\ \bibnamefont
  {Delaney}},\ }\href@noop {} {\bibfield  {journal} {\bibinfo  {journal}
  {Physical Review Letters}\ }\textbf {\bibinfo {volume} {105}},\ \bibinfo
  {pages} {087202} (\bibinfo {year} {2010})}\BibitemShut {NoStop}%
\bibitem [{\citenamefont {Borisov}\ \emph {et~al.}(2007)\citenamefont
  {Borisov}, \citenamefont {Hochstrat}, \citenamefont {Shvartsman},\ and\
  \citenamefont {Kleemann}}]{borisov2007superconducting}%
  \BibitemOpen
  \bibfield  {author} {\bibinfo {author} {\bibfnamefont {P.}~\bibnamefont
  {Borisov}}, \bibinfo {author} {\bibfnamefont {A.}~\bibnamefont {Hochstrat}},
  \bibinfo {author} {\bibfnamefont {V.}~\bibnamefont {Shvartsman}}, \ and\
  \bibinfo {author} {\bibfnamefont {W.}~\bibnamefont {Kleemann}},\ }\href@noop
  {} {\bibfield  {journal} {\bibinfo  {journal} {Review of Scientific
  Instruments}\ }\textbf {\bibinfo {volume} {78}},\ \bibinfo {pages} {106105}
  (\bibinfo {year} {2007})}\BibitemShut {NoStop}%
\bibitem [{\citenamefont {Folen}\ \emph {et~al.}(1961)\citenamefont {Folen},
  \citenamefont {Rado},\ and\ \citenamefont {Stalder}}]{folen1961anisotropy}%
  \BibitemOpen
  \bibfield  {author} {\bibinfo {author} {\bibfnamefont {V.}~\bibnamefont
  {Folen}}, \bibinfo {author} {\bibfnamefont {G.}~\bibnamefont {Rado}}, \ and\
  \bibinfo {author} {\bibfnamefont {E.}~\bibnamefont {Stalder}},\ }\href@noop
  {} {\bibfield  {journal} {\bibinfo  {journal} {Physical Review Letters}\
  }\textbf {\bibinfo {volume} {6}},\ \bibinfo {pages} {607} (\bibinfo {year}
  {1961})}\BibitemShut {NoStop}%
\bibitem [{\citenamefont {Foner}(1963)}]{foner1963high}%
  \BibitemOpen
  \bibfield  {author} {\bibinfo {author} {\bibfnamefont {S.}~\bibnamefont
  {Foner}},\ }\href@noop {} {\bibfield  {journal} {\bibinfo  {journal}
  {Physical Review}\ }\textbf {\bibinfo {volume} {130}},\ \bibinfo {pages}
  {183} (\bibinfo {year} {1963})}\BibitemShut {NoStop}%
\bibitem [{\citenamefont {Abo}\ \emph {et~al.}(2013)\citenamefont {Abo},
  \citenamefont {Hong}, \citenamefont {Park}, \citenamefont {Lee},
  \citenamefont {Lee},\ and\ \citenamefont {Choi}}]{abo2013definition}%
  \BibitemOpen
  \bibfield  {author} {\bibinfo {author} {\bibfnamefont {G.~S.}\ \bibnamefont
  {Abo}}, \bibinfo {author} {\bibfnamefont {Y.-K.}\ \bibnamefont {Hong}},
  \bibinfo {author} {\bibfnamefont {J.}~\bibnamefont {Park}}, \bibinfo {author}
  {\bibfnamefont {J.}~\bibnamefont {Lee}}, \bibinfo {author} {\bibfnamefont
  {W.}~\bibnamefont {Lee}}, \ and\ \bibinfo {author} {\bibfnamefont {B.-C.}\
  \bibnamefont {Choi}},\ }\href@noop {} {\bibfield  {journal} {\bibinfo
  {journal} {IEEE Transactions on Magnetics}\ }\textbf {\bibinfo {volume}
  {49}},\ \bibinfo {pages} {4937} (\bibinfo {year} {2013})}\BibitemShut
  {NoStop}%
\bibitem [{\citenamefont {Landau}\ and\ \citenamefont
  {Lifshitz}(1935)}]{Landau1935dispersion}%
  \BibitemOpen
  \bibfield  {author} {\bibinfo {author} {\bibfnamefont {L.~D.}\ \bibnamefont
  {Landau}}\ and\ \bibinfo {author} {\bibfnamefont {E.}~\bibnamefont
  {Lifshitz}},\ }\href@noop {} {\bibfield  {journal} {\bibinfo  {journal}
  {Phys. Z. Sowjet.}\ }\textbf {\bibinfo {volume} {8}},\ \bibinfo {pages} {153}
  (\bibinfo {year} {1935})}\BibitemShut {NoStop}%
\bibitem [{\citenamefont {Gilbert}(2004)}]{gilbert2004phenomenological}%
  \BibitemOpen
  \bibfield  {author} {\bibinfo {author} {\bibfnamefont {T.~L.}\ \bibnamefont
  {Gilbert}},\ }\href@noop {} {\bibfield  {journal} {\bibinfo  {journal} {IEEE
  Transactions on Magnetics}\ }\textbf {\bibinfo {volume} {40}},\ \bibinfo
  {pages} {3443} (\bibinfo {year} {2004})},\ \bibinfo {note} {{T}he original
  reference,``T. Gilbert, Physical Review 100, 1243 (1955)" is only an abstract
  and its pdf pages are not hosted.}\BibitemShut {Stop}%
\bibitem [{\citenamefont {Aharoni}(2000)}]{aharoni2000introduction}%
  \BibitemOpen
  \bibfield  {author} {\bibinfo {author} {\bibfnamefont {A.}~\bibnamefont
  {Aharoni}},\ }\href@noop {} {\emph {\bibinfo {title} {Introduction to the
  Theory of Ferromagnetism}}},\ Vol.\ \bibinfo {volume} {109}\ (\bibinfo
  {publisher} {Clarendon Press},\ \bibinfo {year} {2000})\BibitemShut {NoStop}%
\bibitem [{\citenamefont {Cullity}\ and\ \citenamefont
  {Graham}(2011)}]{cullity2011introduction}%
  \BibitemOpen
  \bibfield  {author} {\bibinfo {author} {\bibfnamefont {B.~D.}\ \bibnamefont
  {Cullity}}\ and\ \bibinfo {author} {\bibfnamefont {C.~D.}\ \bibnamefont
  {Graham}},\ }\href@noop {} {\emph {\bibinfo {title} {Introduction to magnetic
  materials}}}\ (\bibinfo  {publisher} {John Wiley \& Sons},\ \bibinfo {year}
  {2011})\BibitemShut {NoStop}%
\bibitem [{\citenamefont {Vogel}\ \emph {et~al.}(2006)\citenamefont {Vogel},
  \citenamefont {Moritz},\ and\ \citenamefont
  {Fruchart}}]{vogel2006nucleation}%
  \BibitemOpen
  \bibfield  {author} {\bibinfo {author} {\bibfnamefont {J.}~\bibnamefont
  {Vogel}}, \bibinfo {author} {\bibfnamefont {J.}~\bibnamefont {Moritz}}, \
  and\ \bibinfo {author} {\bibfnamefont {O.}~\bibnamefont {Fruchart}},\
  }\href@noop {} {\bibfield  {journal} {\bibinfo  {journal} {Comptes Rendus
  Physique}\ }\textbf {\bibinfo {volume} {7}},\ \bibinfo {pages} {977}
  (\bibinfo {year} {2006})}\BibitemShut {NoStop}%
\bibitem [{\citenamefont {Ferr{\'e}}(2002)}]{ferre2002}%
  \BibitemOpen
  \bibfield  {author} {\bibinfo {author} {\bibfnamefont {J.}~\bibnamefont
  {Ferr{\'e}}},\ }\enquote {\bibinfo {title} {Dynamics of magnetization
  reversal: From continuous to patterned ferromagnetic films},}\ in\ \href
  {\doibase 10.1007/3-540-40907-6_5} {\emph {\bibinfo {booktitle} {Spin
  Dynamics in Confined Magnetic Structures I}}},\ \bibinfo {editor} {edited by\
  \bibinfo {editor} {\bibfnamefont {B.}~\bibnamefont {Hillebrands}}\ and\
  \bibinfo {editor} {\bibfnamefont {K.}~\bibnamefont {Ounadjela}}}\ (\bibinfo
  {publisher} {Springer Berlin Heidelberg},\ \bibinfo {address} {Berlin,
  Heidelberg},\ \bibinfo {year} {2002})\ pp.\ \bibinfo {pages}
  {127--165}\BibitemShut {NoStop}%
\bibitem [{\citenamefont {Fatuzzo}(1962)}]{fatuzzo1962theoretical}%
  \BibitemOpen
  \bibfield  {author} {\bibinfo {author} {\bibfnamefont {E.}~\bibnamefont
  {Fatuzzo}},\ }\href@noop {} {\bibfield  {journal} {\bibinfo  {journal}
  {Physical Review}\ }\textbf {\bibinfo {volume} {127}},\ \bibinfo {pages}
  {1999} (\bibinfo {year} {1962})}\BibitemShut {NoStop}%
\bibitem [{\citenamefont {Labrune}\ \emph {et~al.}(1989)\citenamefont
  {Labrune}, \citenamefont {Andrieu}, \citenamefont {Rio},\ and\ \citenamefont
  {Bernstein}}]{labrune1989time}%
  \BibitemOpen
  \bibfield  {author} {\bibinfo {author} {\bibfnamefont {M.}~\bibnamefont
  {Labrune}}, \bibinfo {author} {\bibfnamefont {S.}~\bibnamefont {Andrieu}},
  \bibinfo {author} {\bibfnamefont {F.}~\bibnamefont {Rio}}, \ and\ \bibinfo
  {author} {\bibfnamefont {P.}~\bibnamefont {Bernstein}},\ }\href@noop {}
  {\bibfield  {journal} {\bibinfo  {journal} {Journal of Magnetism and Magnetic
  Materials}\ }\textbf {\bibinfo {volume} {80}},\ \bibinfo {pages} {211}
  (\bibinfo {year} {1989})}\BibitemShut {NoStop}%
\bibitem [{\citenamefont {H{\"a}nggi}\ \emph {et~al.}(1990)\citenamefont
  {H{\"a}nggi}, \citenamefont {Talkner},\ and\ \citenamefont
  {Borkovec}}]{hanggi1990reaction}%
  \BibitemOpen
  \bibfield  {author} {\bibinfo {author} {\bibfnamefont {P.}~\bibnamefont
  {H{\"a}nggi}}, \bibinfo {author} {\bibfnamefont {P.}~\bibnamefont {Talkner}},
  \ and\ \bibinfo {author} {\bibfnamefont {M.}~\bibnamefont {Borkovec}},\
  }\href@noop {} {\bibfield  {journal} {\bibinfo  {journal} {Reviews of Modern
  Physics}\ }\textbf {\bibinfo {volume} {62}},\ \bibinfo {pages} {251}
  (\bibinfo {year} {1990})}\BibitemShut {NoStop}%
\bibitem [{\citenamefont {Kim}\ \emph {et~al.}(2014)\citenamefont {Kim},
  \citenamefont {Tserkovnyak},\ and\ \citenamefont
  {Tchernyshyov}}]{kim2014propulsion}%
  \BibitemOpen
  \bibfield  {author} {\bibinfo {author} {\bibfnamefont {S.~K.}\ \bibnamefont
  {Kim}}, \bibinfo {author} {\bibfnamefont {Y.}~\bibnamefont {Tserkovnyak}}, \
  and\ \bibinfo {author} {\bibfnamefont {O.}~\bibnamefont {Tchernyshyov}},\
  }\href@noop {} {\bibfield  {journal} {\bibinfo  {journal} {Physical Review
  B}\ }\textbf {\bibinfo {volume} {90}},\ \bibinfo {pages} {104406} (\bibinfo
  {year} {2014})}\BibitemShut {NoStop}%
\bibitem [{\citenamefont {Belashchenko}\ \emph {et~al.}(2016)\citenamefont
  {Belashchenko}, \citenamefont {Tchernyshyov}, \citenamefont {Kovalev},\ and\
  \citenamefont {Tretiakov}}]{belashchenko2016magnetoelectric}%
  \BibitemOpen
  \bibfield  {author} {\bibinfo {author} {\bibfnamefont {K.~D.}\ \bibnamefont
  {Belashchenko}}, \bibinfo {author} {\bibfnamefont {O.}~\bibnamefont
  {Tchernyshyov}}, \bibinfo {author} {\bibfnamefont {A.~A.}\ \bibnamefont
  {Kovalev}}, \ and\ \bibinfo {author} {\bibfnamefont {O.~A.}\ \bibnamefont
  {Tretiakov}},\ }\href@noop {} {\bibfield  {journal} {\bibinfo  {journal}
  {Applied Physics Letters}\ }\textbf {\bibinfo {volume} {108}},\ \bibinfo
  {pages} {132403} (\bibinfo {year} {2016})}\BibitemShut {NoStop}%
\bibitem [{\citenamefont {Aharoni}(1962)}]{aharoni1962theoretical}%
  \BibitemOpen
  \bibfield  {author} {\bibinfo {author} {\bibfnamefont {A.}~\bibnamefont
  {Aharoni}},\ }\href@noop {} {\bibfield  {journal} {\bibinfo  {journal}
  {Reviews of Modern Physics}\ }\textbf {\bibinfo {volume} {34}},\ \bibinfo
  {pages} {227} (\bibinfo {year} {1962})}\BibitemShut {NoStop}%
\bibitem [{\citenamefont {Krishnan}(2016)}]{krishnan2016fundamentals}%
  \BibitemOpen
  \bibfield  {author} {\bibinfo {author} {\bibfnamefont {K.~M.}\ \bibnamefont
  {Krishnan}},\ }\href@noop {} {\emph {\bibinfo {title} {Fundamentals and
  applications of magnetic materials}}}\ (\bibinfo  {publisher} {Oxford
  University Press},\ \bibinfo {year} {2016})\BibitemShut {NoStop}%
\bibitem [{\citenamefont {Chauve}\ \emph {et~al.}(2000)\citenamefont {Chauve},
  \citenamefont {Giamarchi},\ and\ \citenamefont
  {Le~Doussal}}]{chauve2000creep}%
  \BibitemOpen
  \bibfield  {author} {\bibinfo {author} {\bibfnamefont {P.}~\bibnamefont
  {Chauve}}, \bibinfo {author} {\bibfnamefont {T.}~\bibnamefont {Giamarchi}}, \
  and\ \bibinfo {author} {\bibfnamefont {P.}~\bibnamefont {Le~Doussal}},\
  }\href@noop {} {\bibfield  {journal} {\bibinfo  {journal} {Physical Review
  B}\ }\textbf {\bibinfo {volume} {62}},\ \bibinfo {pages} {6241} (\bibinfo
  {year} {2000})}\BibitemShut {NoStop}%
\bibitem [{\citenamefont {Baltz}\ \emph {et~al.}(2018)\citenamefont {Baltz},
  \citenamefont {Manchon}, \citenamefont {Tsoi}, \citenamefont {Moriyama},
  \citenamefont {Ono},\ and\ \citenamefont
  {Tserkovnyak}}]{baltz2018antiferromagnetic}%
  \BibitemOpen
  \bibfield  {author} {\bibinfo {author} {\bibfnamefont {V.}~\bibnamefont
  {Baltz}}, \bibinfo {author} {\bibfnamefont {A.}~\bibnamefont {Manchon}},
  \bibinfo {author} {\bibfnamefont {M.}~\bibnamefont {Tsoi}}, \bibinfo {author}
  {\bibfnamefont {T.}~\bibnamefont {Moriyama}}, \bibinfo {author}
  {\bibfnamefont {T.}~\bibnamefont {Ono}}, \ and\ \bibinfo {author}
  {\bibfnamefont {Y.}~\bibnamefont {Tserkovnyak}},\ }\href@noop {} {\bibfield
  {journal} {\bibinfo  {journal} {Reviews of Modern Physics}\ }\textbf
  {\bibinfo {volume} {90}},\ \bibinfo {pages} {015005} (\bibinfo {year}
  {2018})}\BibitemShut {NoStop}%
\bibitem [{\citenamefont {Tannous}\ and\ \citenamefont
  {Gieraltowski}(2008)}]{tannous2008stoner}%
  \BibitemOpen
  \bibfield  {author} {\bibinfo {author} {\bibfnamefont {C.}~\bibnamefont
  {Tannous}}\ and\ \bibinfo {author} {\bibfnamefont {J.}~\bibnamefont
  {Gieraltowski}},\ }\href@noop {} {\bibfield  {journal} {\bibinfo  {journal}
  {European Journal of Physics}\ }\textbf {\bibinfo {volume} {29}},\ \bibinfo
  {pages} {475} (\bibinfo {year} {2008})}\BibitemShut {NoStop}%
\bibitem [{\citenamefont {Brown~Jr}(1963)}]{brown1963thermal}%
  \BibitemOpen
  \bibfield  {author} {\bibinfo {author} {\bibfnamefont {W.~F.}\ \bibnamefont
  {Brown~Jr}},\ }\href@noop {} {\bibfield  {journal} {\bibinfo  {journal}
  {Physical Review}\ }\textbf {\bibinfo {volume} {130}},\ \bibinfo {pages}
  {1677} (\bibinfo {year} {1963})}\BibitemShut {NoStop}%
\bibitem [{\citenamefont {Atxitia}\ \emph {et~al.}(2018)\citenamefont
  {Atxitia}, \citenamefont {Birk}, \citenamefont {Selzer},\ and\ \citenamefont
  {Nowak}}]{atxitia2018superparamagnetic}%
  \BibitemOpen
  \bibfield  {author} {\bibinfo {author} {\bibfnamefont {U.}~\bibnamefont
  {Atxitia}}, \bibinfo {author} {\bibfnamefont {T.}~\bibnamefont {Birk}},
  \bibinfo {author} {\bibfnamefont {S.}~\bibnamefont {Selzer}}, \ and\ \bibinfo
  {author} {\bibfnamefont {U.}~\bibnamefont {Nowak}},\ }\href@noop {}
  {\bibfield  {journal} {\bibinfo  {journal} {arXiv preprint arXiv:1808.07665}\
  } (\bibinfo {year} {2018})}\BibitemShut {NoStop}%
\bibitem [{\citenamefont {Mallinson}(2000)}]{mallinson2000damped}%
  \BibitemOpen
  \bibfield  {author} {\bibinfo {author} {\bibfnamefont {J.~C.}\ \bibnamefont
  {Mallinson}},\ }\href@noop {} {\bibfield  {journal} {\bibinfo  {journal}
  {IEEE transactions on magnetics}\ }\textbf {\bibinfo {volume} {36}},\
  \bibinfo {pages} {1976} (\bibinfo {year} {2000})}\BibitemShut {NoStop}%
\bibitem [{\citenamefont {Parthasarathy}\ and\ \citenamefont
  {Rakheja}(2018{\natexlab{a}})}]{ap2018reversal1}%
  \BibitemOpen
  \bibfield  {author} {\bibinfo {author} {\bibfnamefont {A.}~\bibnamefont
  {Parthasarathy}}\ and\ \bibinfo {author} {\bibfnamefont {S.}~\bibnamefont
  {Rakheja}},\ }\href@noop {} {\bibfield  {journal} {\bibinfo  {journal}
  {Journal of Applied Physics}\ }\textbf {\bibinfo {volume} {123}},\ \bibinfo
  {pages} {223901} (\bibinfo {year} {2018}{\natexlab{a}})}\BibitemShut
  {NoStop}%
\bibitem [{\citenamefont {Parthasarathy}\ and\ \citenamefont
  {Rakheja}(2018{\natexlab{b}})}]{ap2018reversal2}%
  \BibitemOpen
  \bibfield  {author} {\bibinfo {author} {\bibfnamefont {A.}~\bibnamefont
  {Parthasarathy}}\ and\ \bibinfo {author} {\bibfnamefont {S.}~\bibnamefont
  {Rakheja}},\ }\href@noop {} {\bibfield  {journal} {\bibinfo  {journal} {IEEE
  Transactions on Magnetics}\ } (\bibinfo {year}
  {2018}{\natexlab{b}})}\BibitemShut {NoStop}%
\bibitem [{\citenamefont {St{\"o}hr}\ and\ \citenamefont
  {Siegmann}(2007)}]{stohr2007magnetism}%
  \BibitemOpen
  \bibfield  {author} {\bibinfo {author} {\bibfnamefont {J.}~\bibnamefont
  {St{\"o}hr}}\ and\ \bibinfo {author} {\bibfnamefont {H.~C.}\ \bibnamefont
  {Siegmann}},\ }\href@noop {} {\emph {\bibinfo {title} {Magnetism: from
  fundamentals to nanoscale dynamics}}},\ Vol.\ \bibinfo {volume} {152}\
  (\bibinfo  {publisher} {Springer Science \& Business Media},\ \bibinfo {year}
  {2007})\BibitemShut {NoStop}%
\bibitem [{\citenamefont {Kirilyuk}\ \emph {et~al.}(2010)\citenamefont
  {Kirilyuk}, \citenamefont {Kimel},\ and\ \citenamefont
  {Rasing}}]{kirilyuk2010ultrafast}%
  \BibitemOpen
  \bibfield  {author} {\bibinfo {author} {\bibfnamefont {A.}~\bibnamefont
  {Kirilyuk}}, \bibinfo {author} {\bibfnamefont {A.~V.}\ \bibnamefont {Kimel}},
  \ and\ \bibinfo {author} {\bibfnamefont {T.}~\bibnamefont {Rasing}},\
  }\href@noop {} {\bibfield  {journal} {\bibinfo  {journal} {Reviews of Modern
  Physics}\ }\textbf {\bibinfo {volume} {82}},\ \bibinfo {pages} {2731}
  (\bibinfo {year} {2010})}\BibitemShut {NoStop}%
\bibitem [{\citenamefont {Samuelsen}\ \emph {et~al.}(1970)\citenamefont
  {Samuelsen}, \citenamefont {Hutchings},\ and\ \citenamefont
  {Shirane}}]{samuelsen1970inelastic}%
  \BibitemOpen
  \bibfield  {author} {\bibinfo {author} {\bibfnamefont {E.}~\bibnamefont
  {Samuelsen}}, \bibinfo {author} {\bibfnamefont {M.}~\bibnamefont
  {Hutchings}}, \ and\ \bibinfo {author} {\bibfnamefont {G.}~\bibnamefont
  {Shirane}},\ }\href@noop {} {\bibfield  {journal} {\bibinfo  {journal}
  {Physica}\ }\textbf {\bibinfo {volume} {48}},\ \bibinfo {pages} {13}
  (\bibinfo {year} {1970})}\BibitemShut {NoStop}%
\bibitem [{\citenamefont {Artman}\ \emph {et~al.}(1965)\citenamefont {Artman},
  \citenamefont {Murphy},\ and\ \citenamefont {Foner}}]{artman1965magnetic}%
  \BibitemOpen
  \bibfield  {author} {\bibinfo {author} {\bibfnamefont {J.}~\bibnamefont
  {Artman}}, \bibinfo {author} {\bibfnamefont {J.}~\bibnamefont {Murphy}}, \
  and\ \bibinfo {author} {\bibfnamefont {S.}~\bibnamefont {Foner}},\
  }\href@noop {} {\bibfield  {journal} {\bibinfo  {journal} {Physical Review}\
  }\textbf {\bibinfo {volume} {138}},\ \bibinfo {pages} {A912} (\bibinfo {year}
  {1965})}\BibitemShut {NoStop}%
\bibitem [{\citenamefont {Wang}\ \emph {et~al.}(2015)\citenamefont {Wang},
  \citenamefont {Du}, \citenamefont {Hammel},\ and\ \citenamefont
  {Yang}}]{wang2015spin}%
  \BibitemOpen
  \bibfield  {author} {\bibinfo {author} {\bibfnamefont {H.}~\bibnamefont
  {Wang}}, \bibinfo {author} {\bibfnamefont {C.}~\bibnamefont {Du}}, \bibinfo
  {author} {\bibfnamefont {P.~C.}\ \bibnamefont {Hammel}}, \ and\ \bibinfo
  {author} {\bibfnamefont {F.}~\bibnamefont {Yang}},\ }\href@noop {} {\bibfield
   {journal} {\bibinfo  {journal} {Physical Review B}\ }\textbf {\bibinfo
  {volume} {91}},\ \bibinfo {pages} {220410} (\bibinfo {year}
  {2015})}\BibitemShut {NoStop}%
\bibitem [{\citenamefont {Brown}\ and\ \citenamefont
  {O'Dell}(1969)}]{brown1969domain}%
  \BibitemOpen
  \bibfield  {author} {\bibinfo {author} {\bibfnamefont {C.}~\bibnamefont
  {Brown}}\ and\ \bibinfo {author} {\bibfnamefont {T.}~\bibnamefont {O'Dell}},\
  }\href@noop {} {\bibfield  {journal} {\bibinfo  {journal} {IEEE Transactions
  on Magnetics}\ }\textbf {\bibinfo {volume} {5}},\ \bibinfo {pages} {964}
  (\bibinfo {year} {1969})}\BibitemShut {NoStop}%
\bibitem [{\citenamefont {Satoh}\ \emph {et~al.}(2007)\citenamefont {Satoh},
  \citenamefont {Van~Aken}, \citenamefont {Duong}, \citenamefont
  {Lottermoser},\ and\ \citenamefont {Fiebig}}]{satoh2007ultrafast}%
  \BibitemOpen
  \bibfield  {author} {\bibinfo {author} {\bibfnamefont {T.}~\bibnamefont
  {Satoh}}, \bibinfo {author} {\bibfnamefont {B.~B.}\ \bibnamefont {Van~Aken}},
  \bibinfo {author} {\bibfnamefont {N.~P.}\ \bibnamefont {Duong}}, \bibinfo
  {author} {\bibfnamefont {T.}~\bibnamefont {Lottermoser}}, \ and\ \bibinfo
  {author} {\bibfnamefont {M.}~\bibnamefont {Fiebig}},\ }\href@noop {}
  {\bibfield  {journal} {\bibinfo  {journal} {Physical Review B}\ }\textbf
  {\bibinfo {volume} {75}},\ \bibinfo {pages} {155406} (\bibinfo {year}
  {2007})}\BibitemShut {NoStop}%
\bibitem [{\citenamefont {Sun}\ \emph {et~al.}(2017)\citenamefont {Sun},
  \citenamefont {Song}, \citenamefont {Rath}, \citenamefont {Street},
  \citenamefont {Echtenkamp}, \citenamefont {Feng}, \citenamefont {Binek},
  \citenamefont {Morgan},\ and\ \citenamefont {Voyles}}]{sun2017local}%
  \BibitemOpen
  \bibfield  {author} {\bibinfo {author} {\bibfnamefont {C.}~\bibnamefont
  {Sun}}, \bibinfo {author} {\bibfnamefont {Z.}~\bibnamefont {Song}}, \bibinfo
  {author} {\bibfnamefont {A.}~\bibnamefont {Rath}}, \bibinfo {author}
  {\bibfnamefont {M.}~\bibnamefont {Street}}, \bibinfo {author} {\bibfnamefont
  {W.}~\bibnamefont {Echtenkamp}}, \bibinfo {author} {\bibfnamefont
  {J.}~\bibnamefont {Feng}}, \bibinfo {author} {\bibfnamefont {C.}~\bibnamefont
  {Binek}}, \bibinfo {author} {\bibfnamefont {D.}~\bibnamefont {Morgan}}, \
  and\ \bibinfo {author} {\bibfnamefont {P.}~\bibnamefont {Voyles}},\ }\href
  {\doibase 10.1002/admi.201700172} {\bibfield  {journal} {\bibinfo  {journal}
  {Advanced Materials Interfaces}\ }\textbf {\bibinfo {volume} {4}},\ \bibinfo
  {pages} {1700172} (\bibinfo {year} {2017})}\BibitemShut {NoStop}%
\bibitem [{\citenamefont {Mu}\ and\ \citenamefont
  {Belashchenko}(2018)}]{mu2018influence}%
  \BibitemOpen
  \bibfield  {author} {\bibinfo {author} {\bibfnamefont {S.}~\bibnamefont
  {Mu}}\ and\ \bibinfo {author} {\bibfnamefont {K.}~\bibnamefont
  {Belashchenko}},\ }\href@noop {} {\bibfield  {journal} {\bibinfo  {journal}
  {arXiv preprint arXiv:1812.04060}\ } (\bibinfo {year} {2018})}\BibitemShut
  {NoStop}%
\bibitem [{Note1()}]{Note1}%
  \BibitemOpen
  \bibinfo {note} {At the time of submission of manuscript, we came across
  experimental works published less than a month before by Shiratsuchi et
  al.,~\cite {shiratsuchi2018observation, shiratsuchi2018antiferromagnetic,
  nguyen2018energy} which corroborate the role of domain wall propagation in ME
  switching of exchange bias. We will assess our theory with their findings in
  a future work.}\BibitemShut {Stop}%
\bibitem [{\citenamefont {Astrov}(1961)}]{astrov1961magnetoelectric}%
  \BibitemOpen
  \bibfield  {author} {\bibinfo {author} {\bibfnamefont {D.}~\bibnamefont
  {Astrov}},\ }\href@noop {} {\bibfield  {journal} {\bibinfo  {journal} {Soviet
  Physics Journal of Experimental and Theoretical Physics}\ }\textbf {\bibinfo
  {volume} {13}},\ \bibinfo {pages} {729} (\bibinfo {year} {1961})}\BibitemShut
  {NoStop}%
\bibitem [{\citenamefont {Borisov}\ \emph {et~al.}(2005)\citenamefont
  {Borisov}, \citenamefont {Hochstrat}, \citenamefont {Chen}, \citenamefont
  {Kleemann},\ and\ \citenamefont {Binek}}]{borisov2005magnetoelectric}%
  \BibitemOpen
  \bibfield  {author} {\bibinfo {author} {\bibfnamefont {P.}~\bibnamefont
  {Borisov}}, \bibinfo {author} {\bibfnamefont {A.}~\bibnamefont {Hochstrat}},
  \bibinfo {author} {\bibfnamefont {X.}~\bibnamefont {Chen}}, \bibinfo {author}
  {\bibfnamefont {W.}~\bibnamefont {Kleemann}}, \ and\ \bibinfo {author}
  {\bibfnamefont {C.}~\bibnamefont {Binek}},\ }\href@noop {} {\bibfield
  {journal} {\bibinfo  {journal} {Physical Review Letters}\ }\textbf {\bibinfo
  {volume} {94}},\ \bibinfo {pages} {117203} (\bibinfo {year}
  {2005})}\BibitemShut {NoStop}%
\bibitem [{\citenamefont {He}\ \emph {et~al.}(2010)\citenamefont {He},
  \citenamefont {Wang}, \citenamefont {Wu}, \citenamefont {Caruso},
  \citenamefont {Vescovo}, \citenamefont {Belashchenko}, \citenamefont
  {Dowben},\ and\ \citenamefont {Binek}}]{he2010robust}%
  \BibitemOpen
  \bibfield  {author} {\bibinfo {author} {\bibfnamefont {X.}~\bibnamefont
  {He}}, \bibinfo {author} {\bibfnamefont {Y.}~\bibnamefont {Wang}}, \bibinfo
  {author} {\bibfnamefont {N.}~\bibnamefont {Wu}}, \bibinfo {author}
  {\bibfnamefont {A.~N.}\ \bibnamefont {Caruso}}, \bibinfo {author}
  {\bibfnamefont {E.}~\bibnamefont {Vescovo}}, \bibinfo {author} {\bibfnamefont
  {K.~D.}\ \bibnamefont {Belashchenko}}, \bibinfo {author} {\bibfnamefont
  {P.~A.}\ \bibnamefont {Dowben}}, \ and\ \bibinfo {author} {\bibfnamefont
  {C.}~\bibnamefont {Binek}},\ }\href@noop {} {\bibfield  {journal} {\bibinfo
  {journal} {Nature Materials}\ }\textbf {\bibinfo {volume} {9}},\ \bibinfo
  {pages} {579} (\bibinfo {year} {2010})}\BibitemShut {NoStop}%
\bibitem [{\citenamefont {Belashchenko}(2010)}]{belashchenko2010equilibrium}%
  \BibitemOpen
  \bibfield  {author} {\bibinfo {author} {\bibfnamefont {K.~D.}\ \bibnamefont
  {Belashchenko}},\ }\href@noop {} {\bibfield  {journal} {\bibinfo  {journal}
  {Physical Review Letters}\ }\textbf {\bibinfo {volume} {105}},\ \bibinfo
  {pages} {147204} (\bibinfo {year} {2010})}\BibitemShut {NoStop}%
\bibitem [{\citenamefont {Mu}\ \emph {et~al.}(2013)\citenamefont {Mu},
  \citenamefont {Wysocki},\ and\ \citenamefont {Belashchenko}}]{mu2013effect}%
  \BibitemOpen
  \bibfield  {author} {\bibinfo {author} {\bibfnamefont {S.}~\bibnamefont
  {Mu}}, \bibinfo {author} {\bibfnamefont {A.~L.}\ \bibnamefont {Wysocki}}, \
  and\ \bibinfo {author} {\bibfnamefont {K.~D.}\ \bibnamefont {Belashchenko}},\
  }\href@noop {} {\bibfield  {journal} {\bibinfo  {journal} {Physical Review
  B}\ }\textbf {\bibinfo {volume} {87}},\ \bibinfo {pages} {054435} (\bibinfo
  {year} {2013})}\BibitemShut {NoStop}%
\bibitem [{\citenamefont {Street}\ \emph {et~al.}(2014)\citenamefont {Street},
  \citenamefont {Echtenkamp}, \citenamefont {Komesu}, \citenamefont {Cao},
  \citenamefont {Dowben},\ and\ \citenamefont {Binek}}]{street2014increasing}%
  \BibitemOpen
  \bibfield  {author} {\bibinfo {author} {\bibfnamefont {M.}~\bibnamefont
  {Street}}, \bibinfo {author} {\bibfnamefont {W.}~\bibnamefont {Echtenkamp}},
  \bibinfo {author} {\bibfnamefont {T.}~\bibnamefont {Komesu}}, \bibinfo
  {author} {\bibfnamefont {S.}~\bibnamefont {Cao}}, \bibinfo {author}
  {\bibfnamefont {P.~A.}\ \bibnamefont {Dowben}}, \ and\ \bibinfo {author}
  {\bibfnamefont {C.}~\bibnamefont {Binek}},\ }\href@noop {} {\bibfield
  {journal} {\bibinfo  {journal} {Applied Physics Letters}\ }\textbf {\bibinfo
  {volume} {104}},\ \bibinfo {pages} {222402} (\bibinfo {year}
  {2014})}\BibitemShut {NoStop}%
\bibitem [{\citenamefont {Ashida}\ \emph {et~al.}(2015)\citenamefont {Ashida},
  \citenamefont {Oida}, \citenamefont {Shimomura}, \citenamefont {Nozaki},
  \citenamefont {Shibata},\ and\ \citenamefont
  {Sahashi}}]{ashida2015isothermal}%
  \BibitemOpen
  \bibfield  {author} {\bibinfo {author} {\bibfnamefont {T.}~\bibnamefont
  {Ashida}}, \bibinfo {author} {\bibfnamefont {M.}~\bibnamefont {Oida}},
  \bibinfo {author} {\bibfnamefont {N.}~\bibnamefont {Shimomura}}, \bibinfo
  {author} {\bibfnamefont {T.}~\bibnamefont {Nozaki}}, \bibinfo {author}
  {\bibfnamefont {T.}~\bibnamefont {Shibata}}, \ and\ \bibinfo {author}
  {\bibfnamefont {M.}~\bibnamefont {Sahashi}},\ }\href@noop {} {\bibfield
  {journal} {\bibinfo  {journal} {Applied Physics Letters}\ }\textbf {\bibinfo
  {volume} {106}},\ \bibinfo {pages} {132407} (\bibinfo {year}
  {2015})}\BibitemShut {NoStop}%
\bibitem [{\citenamefont {Kosub}\ \emph {et~al.}(2015)\citenamefont {Kosub},
  \citenamefont {Kopte}, \citenamefont {Radu}, \citenamefont {Schmidt},\ and\
  \citenamefont {Makarov}}]{kosub2015all}%
  \BibitemOpen
  \bibfield  {author} {\bibinfo {author} {\bibfnamefont {T.}~\bibnamefont
  {Kosub}}, \bibinfo {author} {\bibfnamefont {M.}~\bibnamefont {Kopte}},
  \bibinfo {author} {\bibfnamefont {F.}~\bibnamefont {Radu}}, \bibinfo {author}
  {\bibfnamefont {O.~G.}\ \bibnamefont {Schmidt}}, \ and\ \bibinfo {author}
  {\bibfnamefont {D.}~\bibnamefont {Makarov}},\ }\href@noop {} {\bibfield
  {journal} {\bibinfo  {journal} {Physical Review Letters}\ }\textbf {\bibinfo
  {volume} {115}},\ \bibinfo {pages} {097201} (\bibinfo {year}
  {2015})}\BibitemShut {NoStop}%
\bibitem [{\citenamefont {Rado}(1961)}]{rado1961mechanism}%
  \BibitemOpen
  \bibfield  {author} {\bibinfo {author} {\bibfnamefont {G.~T.}\ \bibnamefont
  {Rado}},\ }\href@noop {} {\bibfield  {journal} {\bibinfo  {journal} {Physical
  Review Letters}\ }\textbf {\bibinfo {volume} {6}},\ \bibinfo {pages} {609}
  (\bibinfo {year} {1961})}\BibitemShut {NoStop}%
\bibitem [{\citenamefont {Date}\ \emph {et~al.}(1961)\citenamefont {Date},
  \citenamefont {Kanamori},\ and\ \citenamefont {Tachiki}}]{date1961origin}%
  \BibitemOpen
  \bibfield  {author} {\bibinfo {author} {\bibfnamefont {M.}~\bibnamefont
  {Date}}, \bibinfo {author} {\bibfnamefont {J.}~\bibnamefont {Kanamori}}, \
  and\ \bibinfo {author} {\bibfnamefont {M.}~\bibnamefont {Tachiki}},\
  }\href@noop {} {\bibfield  {journal} {\bibinfo  {journal} {Journal of the
  Physical Society of Japan}\ }\textbf {\bibinfo {volume} {16}},\ \bibinfo
  {pages} {2589} (\bibinfo {year} {1961})}\BibitemShut {NoStop}%
\bibitem [{\citenamefont {Hornreich}\ and\ \citenamefont
  {Shtrikman}(1967)}]{hornreich1967statistical}%
  \BibitemOpen
  \bibfield  {author} {\bibinfo {author} {\bibfnamefont {R.}~\bibnamefont
  {Hornreich}}\ and\ \bibinfo {author} {\bibfnamefont {S.}~\bibnamefont
  {Shtrikman}},\ }\href@noop {} {\bibfield  {journal} {\bibinfo  {journal}
  {Physical Review}\ }\textbf {\bibinfo {volume} {161}},\ \bibinfo {pages}
  {506} (\bibinfo {year} {1967})}\BibitemShut {NoStop}%
\bibitem [{\citenamefont {Bousquet}\ \emph {et~al.}(2011)\citenamefont
  {Bousquet}, \citenamefont {Spaldin},\ and\ \citenamefont
  {Delaney}}]{bousquet2011unexpectedly}%
  \BibitemOpen
  \bibfield  {author} {\bibinfo {author} {\bibfnamefont {E.}~\bibnamefont
  {Bousquet}}, \bibinfo {author} {\bibfnamefont {N.~A.}\ \bibnamefont
  {Spaldin}}, \ and\ \bibinfo {author} {\bibfnamefont {K.~T.}\ \bibnamefont
  {Delaney}},\ }\href@noop {} {\bibfield  {journal} {\bibinfo  {journal}
  {Physical Review Letters}\ }\textbf {\bibinfo {volume} {106}},\ \bibinfo
  {pages} {107202} (\bibinfo {year} {2011})}\BibitemShut {NoStop}%
\bibitem [{\citenamefont {Malashevich}\ \emph {et~al.}(2012)\citenamefont
  {Malashevich}, \citenamefont {Coh}, \citenamefont {Souza},\ and\
  \citenamefont {Vanderbilt}}]{malashevich2012full}%
  \BibitemOpen
  \bibfield  {author} {\bibinfo {author} {\bibfnamefont {A.}~\bibnamefont
  {Malashevich}}, \bibinfo {author} {\bibfnamefont {S.}~\bibnamefont {Coh}},
  \bibinfo {author} {\bibfnamefont {I.}~\bibnamefont {Souza}}, \ and\ \bibinfo
  {author} {\bibfnamefont {D.}~\bibnamefont {Vanderbilt}},\ }\href@noop {}
  {\bibfield  {journal} {\bibinfo  {journal} {Physical Review B}\ }\textbf
  {\bibinfo {volume} {86}},\ \bibinfo {pages} {094430} (\bibinfo {year}
  {2012})}\BibitemShut {NoStop}%
\bibitem [{\citenamefont {Andreev}(1996)}]{andreev1996macroscopic}%
  \BibitemOpen
  \bibfield  {author} {\bibinfo {author} {\bibfnamefont {A.}~\bibnamefont
  {Andreev}},\ }\href@noop {} {\bibfield  {journal} {\bibinfo  {journal}
  {Journal of Experimental and Theoretical Physics Letters}\ }\textbf {\bibinfo
  {volume} {63}},\ \bibinfo {pages} {758} (\bibinfo {year} {1996})}\BibitemShut
  {NoStop}%
\bibitem [{Vil()}]{Villars2016}%
  \BibitemOpen
  \href {https://materials.springer.com/isp/crystallographic/docs/sd_1715495}
  {\enquote {\bibinfo {title} {Cr\textsubscript{2}{O}\textsubscript{3} crystal
  structure: Datasheet from ``pauling file multinaries edition -- 2012'' in
  springermaterials},}\ }\BibitemShut {NoStop}%
\bibitem [{\citenamefont {Shiratsuchi}\ \emph
  {et~al.}(2018{\natexlab{a}})\citenamefont {Shiratsuchi}, \citenamefont
  {Watanabe}, \citenamefont {Yoshida}, \citenamefont {Kishida}, \citenamefont
  {Nakatani}, \citenamefont {Kotani}, \citenamefont {Toyoki},\ and\
  \citenamefont {Nakamura}}]{shiratsuchi2018observation}%
  \BibitemOpen
  \bibfield  {author} {\bibinfo {author} {\bibfnamefont {Y.}~\bibnamefont
  {Shiratsuchi}}, \bibinfo {author} {\bibfnamefont {S.}~\bibnamefont
  {Watanabe}}, \bibinfo {author} {\bibfnamefont {H.}~\bibnamefont {Yoshida}},
  \bibinfo {author} {\bibfnamefont {N.}~\bibnamefont {Kishida}}, \bibinfo
  {author} {\bibfnamefont {R.}~\bibnamefont {Nakatani}}, \bibinfo {author}
  {\bibfnamefont {Y.}~\bibnamefont {Kotani}}, \bibinfo {author} {\bibfnamefont
  {K.}~\bibnamefont {Toyoki}}, \ and\ \bibinfo {author} {\bibfnamefont
  {T.}~\bibnamefont {Nakamura}},\ }\href@noop {} {\bibfield  {journal}
  {\bibinfo  {journal} {Applied Physics Letters}\ }\textbf {\bibinfo {volume}
  {113}},\ \bibinfo {pages} {242404} (\bibinfo {year}
  {2018}{\natexlab{a}})}\BibitemShut {NoStop}%
\bibitem [{\citenamefont {Shiratsuchi}\ \emph
  {et~al.}(2018{\natexlab{b}})\citenamefont {Shiratsuchi}, \citenamefont
  {Yoshida}, \citenamefont {Kotani}, \citenamefont {Toyoki}, \citenamefont
  {Nguyen}, \citenamefont {Nakamura},\ and\ \citenamefont
  {Nakatani}}]{shiratsuchi2018antiferromagnetic}%
  \BibitemOpen
  \bibfield  {author} {\bibinfo {author} {\bibfnamefont {Y.}~\bibnamefont
  {Shiratsuchi}}, \bibinfo {author} {\bibfnamefont {H.}~\bibnamefont
  {Yoshida}}, \bibinfo {author} {\bibfnamefont {Y.}~\bibnamefont {Kotani}},
  \bibinfo {author} {\bibfnamefont {K.}~\bibnamefont {Toyoki}}, \bibinfo
  {author} {\bibfnamefont {T.~V.~A.}\ \bibnamefont {Nguyen}}, \bibinfo {author}
  {\bibfnamefont {T.}~\bibnamefont {Nakamura}}, \ and\ \bibinfo {author}
  {\bibfnamefont {R.}~\bibnamefont {Nakatani}},\ }\href@noop {} {\bibfield
  {journal} {\bibinfo  {journal} {APL Materials}\ }\textbf {\bibinfo {volume}
  {6}},\ \bibinfo {pages} {121104} (\bibinfo {year}
  {2018}{\natexlab{b}})}\BibitemShut {NoStop}%
\bibitem [{\citenamefont {Nguyen}\ \emph {et~al.}(2018)\citenamefont {Nguyen},
  \citenamefont {Shiratsuchi}, \citenamefont {Yonemura}, \citenamefont
  {Shibata},\ and\ \citenamefont {Nakatani}}]{nguyen2018energy}%
  \BibitemOpen
  \bibfield  {author} {\bibinfo {author} {\bibfnamefont {T.~V.~A.}\
  \bibnamefont {Nguyen}}, \bibinfo {author} {\bibfnamefont {Y.}~\bibnamefont
  {Shiratsuchi}}, \bibinfo {author} {\bibfnamefont {S.}~\bibnamefont
  {Yonemura}}, \bibinfo {author} {\bibfnamefont {T.}~\bibnamefont {Shibata}}, \
  and\ \bibinfo {author} {\bibfnamefont {R.}~\bibnamefont {Nakatani}},\
  }\href@noop {} {\bibfield  {journal} {\bibinfo  {journal} {Journal of Applied
  Physics}\ }\textbf {\bibinfo {volume} {124}},\ \bibinfo {pages} {233902}
  (\bibinfo {year} {2018})}\BibitemShut {NoStop}%
\end{thebibliography}%

\end{document}